\documentclass[12pt,preprint]{aastex}

\usepackage{graphicx}
\usepackage{gensymb}
\usepackage{lscape}

\slugcomment{Accepted to ApJ.}

\shorttitle{Molecular Gas in NGC1068}
\shortauthors{M\"uller S\'anchez et al.}

\begin{document}


\title{Molecular gas streamers feeding and obscuring the active nucleus of NGC1068\footnotemark[1]}


\author{F. M\"uller S\'anchez$^1$, R.~I. Davies$^1$, R. Genzel$^{1,2}$, 
        L.~J. Tacconi$^1$, F. Eisenhauer$^1$, E.~K.~S. Hicks$^1$, S. Friedrich$^1$, A. Sternberg$^3$}
\affil{$^1$ Max-Planck-Institut f\"ur extraterrestrische Physik,
        Giessenbachstrasse, Postfach 1312, D-85741 Garching}

\affil{$^2$ Department of Physics, 366 Le~Conte Hall, University of 
        California, Berkeley, CA, 94720-7300, United States}

\affil{$^3$ School of Physics and Astronomy, Tel Aviv University, Tel Aviv 69978, Israel}

\footnotetext[1]{Based on observations at the European Southern Observatory VLT (076.B-0098).}

\begin{abstract}
We report the first direct observations of neutral, molecular gas streaming in the nucleus of NGC1068 on scales of $<30$ pc using SINFONI near-infrared integral field spectroscopy. At a resolution of $0.075\arcsec$, the flux map of 2.12 $\mu$m 1--0 S(1) molecular hydrogen emission around the nucleus in the central arcsec reveals two prominent linear structures leading to the AGN from the north and south. The kinematics of the gas in these features are dominated by non-circular motions and indicate that material is streaming towards the nucleus on highly elliptical or parabolic trajectories whose orientations are compatible with that of the disk plane of the galaxy. We interpret the data as evidence for fueling of gas to the central region. 
The radial transport rate from $\sim30$ pc to a few parsec from the nucleus is $\sim$15 M$_\sun$ yr$^{-1}$. One of the infalling clouds lies directly in front of the central engine. We interpret it as a tidally disrupted streamer that forms the optically thick outerpart of an amorphous clumpy molecular/dusty structure which contributes to the nuclear obscuration.   

\end{abstract}


\section{Introduction}

A detailed description of the distribution and kinematics of the molecular gas in the central region of Seyfert galaxies is crucial for understanding the fueling of the nucleus and the role of gas in obscuring the active galactic nucleus (AGN). NGC1068, at a distance of 14.4 Mpc \citep{bland97}, is the brightest Seyfert 2 galaxy and therefore one of the best candidates for direct investigations of the morphology and dynamics of the molecular gas neighboring an active nucleus. 
The idea of a rotating molecular/dusty torus surrounding a supermassive black hole emerged from the interpretation given by 
\citet{miller83} of optical spectropolarimetry of precisely this galaxy, 
which revealed scattered type--I emission from the obscured broad-line region (BLR). 
Since then, several near- and mid-IR observations have in fact found and begun to elucidate the physical conditions of the concentration of molecular gas and dust in the nucleus of NGC1068 (near-IR: Young et al. 1996; Marco et al. 1997; Rouan et al. 1998, 2004; Alloin et al. 2001; Galliano \& Alloin 2002; Galliano et al. 2003; Weigelt et al. 2004; Wittkowski et al. 2004; Gratadour et al. 2005, 2006; mid-IR: Bock et al. 2000, Tomono et al. 2001, 2006; Jaffe et al. 2004; Galliano et al. 2005; Mason et al. 2006; Poncelet et al. 2006, 2007), all favouring indirectly the existence of a compact molecular/dusty torus but failing to obtain a clear image of it. Alternatively, by means of mid-IR observations over the central 140 pc (2$\arcsec$), \citet{cameron93} proposed a model in which the bulk of the molecular gas and dust is located at large distances (several tens of parsecs) from the AGN. Line-of-sight attenuation of the BLR in this case would be a mere consequence of one or more intervening molecular clouds. More recently, \citet{jaffe07} analyzed new interferometric mid-IR observations of NGC1068 and found that the fitted Gaussian components to the $(u,v)$ plane of the central 10 $\mu$m source resemble a disk similar to the H$_2$O masers \citep{greenhill96}. Both, the dust and H$_2$O maser disks, appear to be oriented neither perpendicular to nor aligned with the radio jet. However, one needs to be cautious when interpreting all of these observations. As they are measurements of the continuum emission, the inferred gas/dust distributions are strongly dependent on temperature, tracing in fact radiation of matter at a given temperature rather than spatial gas/dust distributions. 

The 2.122 $\mu$m H$_2$ rovibrational $\nu=$1--0 S(1) emission line probes hot ($\geq10^3$ K) and moderately dense ($\geq10^3$ cm$^{-3}$) molecular gas and as such, may be an excellent tracer of gas in the nuclear region. Its spatial distribution can expose the potential presence of a molecular/dusty torus, and also provide physical information on fueling or feedback mechanisms. 
Imaging spectroscopy of this line \citep{rotaciuc91, blietz94} has indicated the presence of significant amounts of hot, dense, circumnuclear molecular gas extending over the central 4 arcseconds which is associated with the narrow-line clouds. The nuclear region shows a strong peak almost $1\arcsec$ east of the nucleus and a weaker one $\sim1\arcsec$ to the west. Furthermore, the H$_2$ emission at a few arcseconds from the nucleus is more extended along the major axis of the bar \citep{davies98}.  
More recently, \citet{galliano02} obtained 2D spectroscopic observations of the warm molecular gas in the central $4\arcsec\times4\arcsec$ of the galaxy. They did not detect H$_2$ emission at the location of the $K-$band continuum core (which is known already to be coincident with the central engine at these scales). Instead they confirmed the two main regions of H$_2$ emission at about $1\arcsec$ east and west of the nucleus along a PA=$90\degree$, and a region with complex line profiles at $\sim50$ pc north of the AGN. They interpreted the observed H$_2$ emission at these scales in terms of a warped disk consistent with the interpretation given by \citet{schinnerer00} to millimeter interferometric maps of $^{12}$CO(2--1) emission. It is now apparent from recent SINFONI H$_2$ 1--0 S(1) data at these scales that simple warped 
disk models of the molecular gas \citep{schinnerer00, baker00} cannot
account for the fantastic variety and detail in the morphological and kinematical structure (see Davies et al. 2006), in particular since there is a considerable amount of gas inside the inner edge of the 100 pc ring.

In order to analyze in detail the physical conditions of the warm molecular gas in the central region of an active nucleus, we have obtained adaptive optics assisted SINFONI integral field spectroscopic data to peer deeply into the energetic core of NGC1068. 
In this paper we focus on the observations of the molecular gas emission 
in the central arcsecond of the galaxy done with the smallest spatial scale of the instrument, resulting in a resolution of 75 milli-arcseconds, approximately 7 times better than previous observations. We refer to \citet{mueller08} for the analysis of the stars, molecular and ionized gas in the central $3\arcsec$ of the galaxy corresponding to the rest of the results from the SINFONI observations of NGC1068. 

\section{Observations} \label{observations}

The data presented here were obtained during 2005--2006 using the adaptive optics assisted near-infrared integral field spectrograph SINFONI \citep{bonnet04, eisenhau03} on the VLT UT4. SINFONI delivers spectra simultaneously over a contiguous two-dimensional field of $64\times32$ pixels. The data were taken in the $H+K-$bands using two of the three possible spatial pixel scales of the instrument: $0.0125\arcsec\times0.025\arcsec$ and $0.05\arcsec\times0.1\arcsec$. 

The $0.0125\arcsec\times0.025\arcsec$ SINFONI data were taken in two sets of 2 and 1.5h integrations on the nights of 21 October 2005 and 27 November 2006, both using the galaxy nucleus as reference for the MACAO adaptive optics system. In both nights the atmospheric conditions were excellent (optical seeing of the two nights was $\sim 0.7\arcsec$ and $\sim 0.5\arcsec$ respectively), which allows us to obtain a resolution of 0.075$\arcsec$ FWHM as measured from the spatially marginally resolved non-stellar continuum in $K-$band. The $0.05\arcsec\times0.1\arcsec$ data were obtained on 6 October 2005. Once more, the AO module was able to correct on the nucleus of NGC1068 in seeing of $0.63\arcsec$, to reach $\sim0.1\arcsec$ spatial resolution. In this case, a total of 4 sky and 8 on-source exposures of 50 sec each, were combined to make the final data cube. For the $0.0125\arcsec\times0.025\arcsec$ and $0.05\arcsec\times0.1\arcsec$ datasets, the spectrograph was able, in a single shot, to obtain spectra covering the whole $H$ and $K-$bands (approximately $1.45-2.45$ $\mu$m) at a spectral resolution FWHM of 125 km s$^{-1}$ (R$\sim$2400) for each pixel in the $0.8\arcsec\times0.8\arcsec$ and $3.2\arcsec\times3.2\arcsec$ field of view.

The data were reduced using the SINFONI custom reduction package SPRED \citep{abuter05}. This performs all the usual steps needed to reduce near infrared spectra, but with the additional routines for reconstructing the data cube. Flux calibration was performed simultaneously in $H$ and $K-$bands using a G2V star HD20339 ($H$ = 6.45, $K$=6.41), which yielded in a 1$\arcsec$ aperture a $K-$band magnitude for NGC1068 of 7.73 and $H-$band magnitude of 10.18. The values we deduced are consistent, at similar and smaller apertures, with values found in previous studies \citep{rouan98, gratadour06, prieto05}.  
In addition, flux calibration was cross-checked with VLT NACO 
data in $2-4\arcsec$ apertures \citep{mueller08}. The NACO data was further cross-checked in larger $5-10\arcsec$ apertures using 2MASS data. Agreement between the different data sources was consistent to 15\%.

\section{Distribution, kinematics and physical properties of the molecular gas} \label{results}

\subsection{Molecular gas morphology} \label{morphology}

\begin{figure}
\epsscale{.99}
\plotone{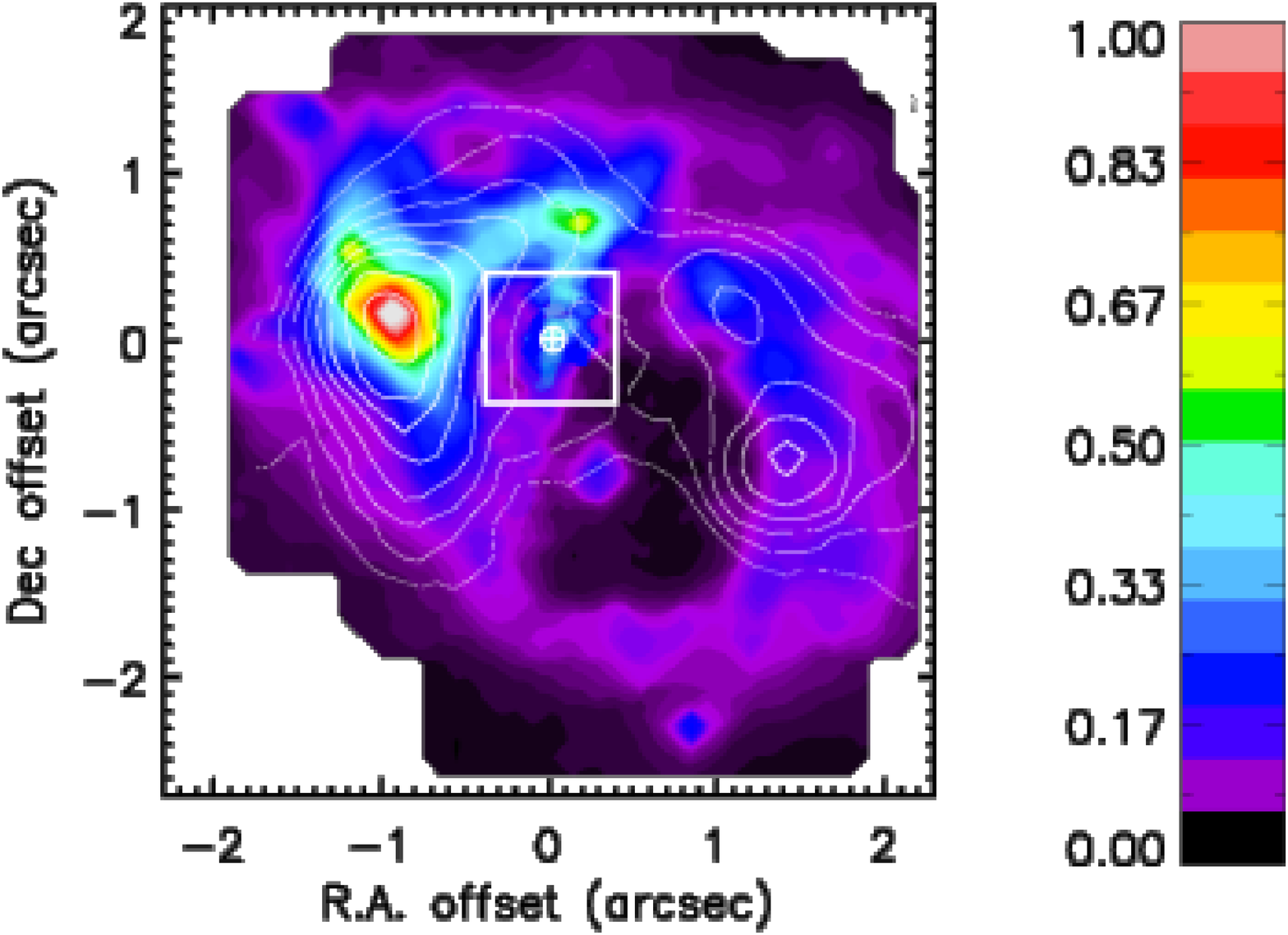}
\caption{SINFONI 2.12 $\mu$m H$_2$ 1--0 S(1) flux map of the central region of NGC1068 at a resolution of $0.1\arcsec$. A contour plot of the $^{12}$CO(2--1) emission map from \citet{schinnerer00} is overlaid on the H$_2$ flux map. 
The scale is given in arcsec ($1\arcsec=70$ pc). The position of the nucleus as defined by the near-IR non-stellar peak is represented by a crossed circle. In addition to the previously known prominent gas concentrations $\sim1\arcsec$ east and west of the nucleus, the morphology reveals several new distinct structures. One of them is a complete ring centered $\sim0.6\arcsec$ SW of the nucleus. The other is a linear structure apparently emanating from the circum-nuclear ring and connecting to the AGN along a PA of $\sim-14\degree$. This new linear streamer is the focus of the discussion in this paper. The central square emphasize the location of this linear structure and denotes the field of the small scale observations shown in Figures 2 -- 8. \label{fig1}}
\end{figure}

The molecular hydrogen emission at scales of a few arcseconds from the nucleus has been mapped previously reaching spatial resolutions down to $\approx 0.5\arcsec$ (Galliano \& Alloin 2002). 
Our SINFONI data at these scales reach $\sim 0.1\arcsec$ resolution and reveal a complex distribution of the gas which has not been entirely observed before (Fig.~\ref{fig1}). 
The new data resolve the previously studied H$_2$ knots and show the presence of an off-center ($\Delta\,r=0.6\arcsec$ SW from the AGN) ring of molecular gas (see Fig.~\ref{fig9}), as well as apparently linear gas streamers leading from the ring NNW and SSE to the center. 
In addition, the overlay of our H$_2$ flux map and the $^{12}$CO(2--1) map from \citet{schinnerer00} at these scales (Fig.~\ref{fig1}) shows a very good correlation, confirming that gas distributions can be traced either by the 2.122 $\mu$m H$_2$ 1--0 S(1) emission line (hot gas) or the $^{12}$CO(2--1) emission (cold gas).  
We discuss briefly the morphology of the ring in Section~\ref{connection} but refer to \citet{mueller08} for more analysis and discussion of these results. In the following we will concentrate on the central arcsecond (white box in Fig.~\ref{fig1}) and provide a few remarks on the connection of the linear streamers to the circumnuclear environment. 

\begin{figure}
\epsscale{.99}
\plotone{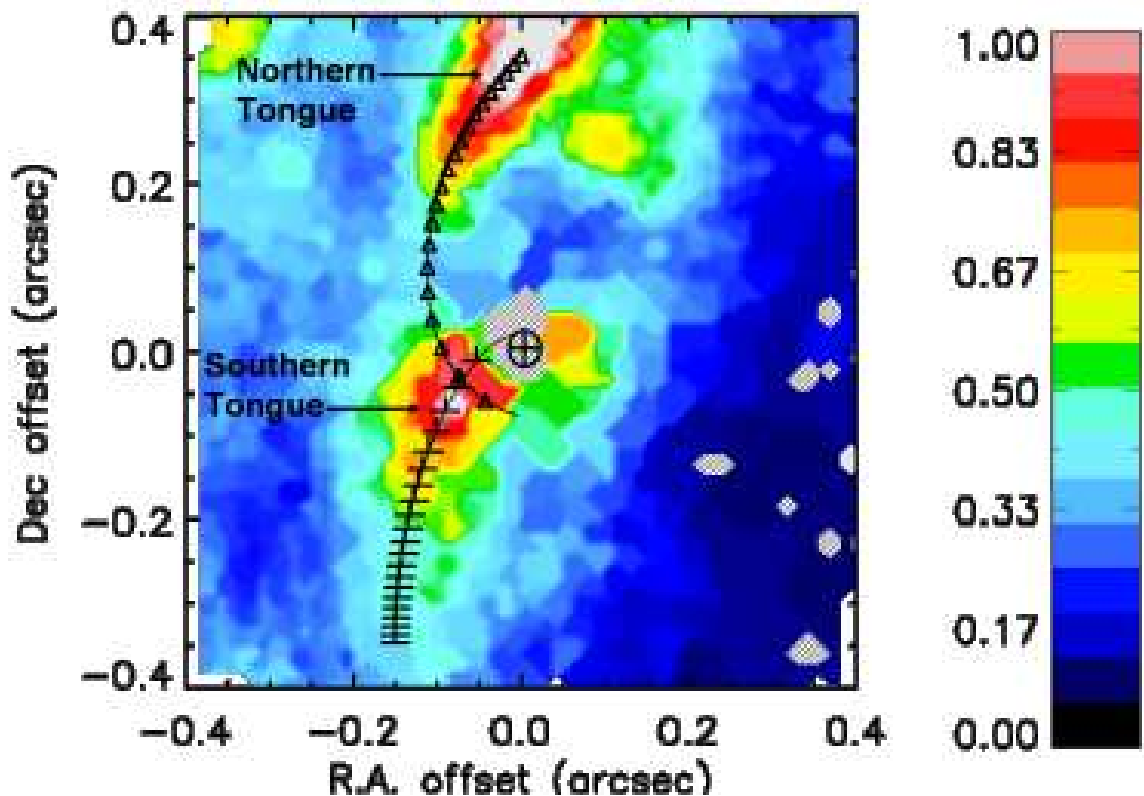}
\caption{Flux map of the H$_2$ 1--0 S(1) emission in the central $\pm0.4\arcsec$ of NGC1068 with a resolution of 0.075$\arcsec$ ($\sim5$ pc). The peak of the non-stellar continuum is represented by a crossed circle. The image is binned using Voronoi tessellations \citep{voronoi}. The line properties could not be extracted in the very central region around the AGN due to the strong continuum emission. The bins where the line properties could not be extracted were masked out and correspond to the diagonal patterned regions. The open triangles show the projected trajectory of the northern concentration of gas (the northern tongue, Orbit NT2). The half-crosses show the past trajectory of the gas which is currently located in front of the AGN (the southern tongue, Orbit ST2).\label{fig2}}
\end{figure}

The morphology of the H$_2$ 1--0S(1) emission in the central $0.8\arcsec\times0.8\arcsec$ of NGC1068 with a resolution of 0.075$\arcsec$ ($\sim$5 pc) is presented in Fig.~\ref{fig2}. The peak of the 2.1 $\mu$m non-stellar continuum is located at the origin of the image and is represented by a crossed circle. At these scales this can be identified as the position of the central engine \citep{galliano03}. At larger scales, a linear structure leading to the AGN from the NNW and SSE is the only noticeable feature in this whole region (see Fig.~\ref{fig1}). On sub-arcsecond scales near the AGN, this linear structure exhibits two prominent regions of H$_2$ emission, one overlapping the center of the non-stellar continuum, and another one located $\sim$0.35$\arcsec$ north of this point.

\begin{figure}
\epsscale{.99}
\plotone{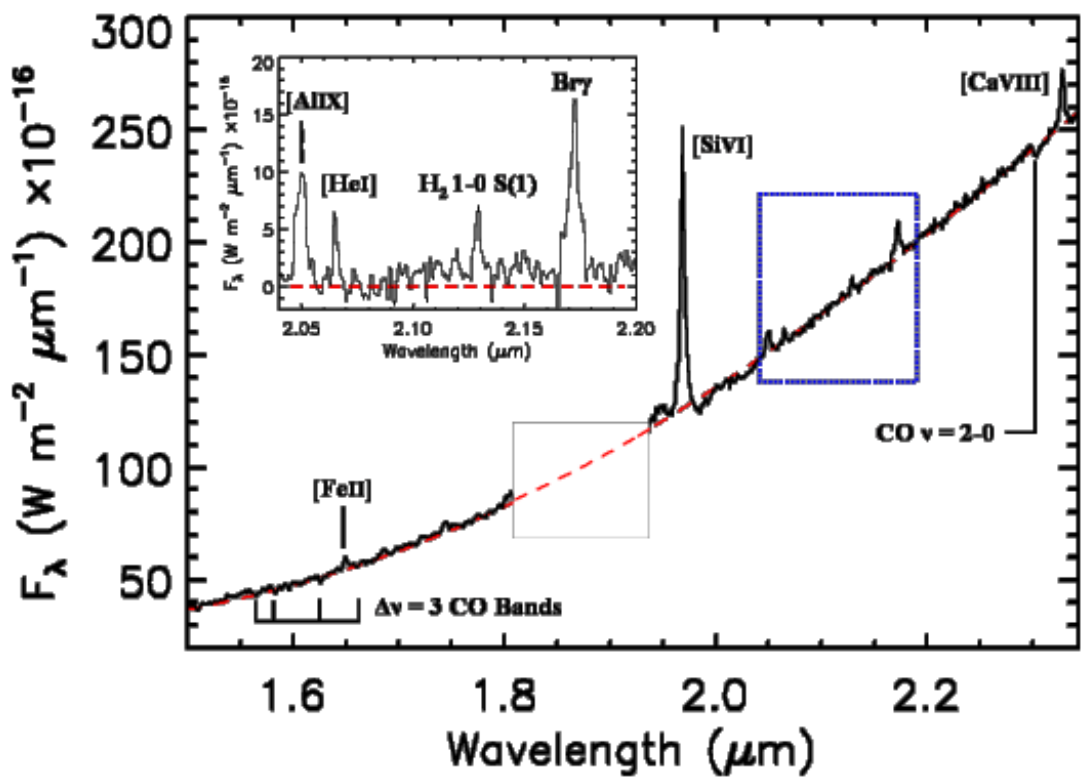}
\caption{Integrated spectrum of the nucleus of NGC1068 in the wavelength range between 1.5 and 2.35 $\mu$m ($H+K-$bands). The individual spectra were added over a $0.3\arcsec\times0.1\arcsec$ rectangular aperture centered at the position of the nucleus. The dashed line represents the best fitting curve to the continumm emission which corresponds to a black body radiating at 700 K. The dotted square emphazises the region between 2.05 and 2.20 $\mu$m, the wavelength range where the H$_2$ 1--0 S(1) emission line can be found. This region of the electromagnetic spectrum is plotted in the small panel top left. The continuum has been substracted in this panel. The H$_2$ 1--0 S(1) emission line is clearly seen. Other relevant emission lines are indicated in the figure.\label{fig3}}
\end{figure}

Thanks to improved spatial resolution and sensitivity, molecular gas is now detected closer and closer to the AGN, in contrast to previous observations \citep{galliano02}. In Fig.~\ref{fig3}, we show a spectrum of NGC1068 at a spectral resolution of R$\sim$2400, integrated over a $0.3\arcsec\times0.1\arcsec$ rectangular aperture centered at the position of the nucleus; although the continuum is quite strong in this region, the H$_2$ 1--0 S(1) line is clearly seen. We will refer to this nuclear concentration of gas as the \textbf{\textit{southern tongue}}. It has a major axis of $\sim$17 pc and a minor axis of $\sim$7 pc and is elongated to the south following the SSE extension of the linear structure. Its major axis has a position angle of about 120$\degree$, similar to that of the H$_2$O maser disk \citep{greenhill96} and the 10 $\mu$m dust emission \citep{jaf04, poncelet07} on a scale of 20 mas. These three features exhibit the same misallignment with the major axis of the nuclear 5 GHz radio continuum \citep{gallimore04}. The H$_2$ 1--0 S(1) luminosity of the southern tongue is $1.5\times10^{-18}$ W m$^{-2}$ ($\sim 1\times10^4\, \mathrm{L_\sun}$). 

In addition to the nuclear component designated as the southern tongue, we found another area of high luminosity in the northern part of the H$_2$ intensity map (see Fig.~\ref{fig2}). In fact, this is the brightest region of the linear structure ($\sim 1.6\times10^4\, \mathrm{L_\sun}$) which extends up to $\sim0.7\arcsec$ NNW of the central engine along a PA of $-14\degree$ and apparently connects to a prominent knot in the circumnuclear ring (Fig.~\ref{fig1}). We will designate this extension of the linear structure as the \textbf{\textit{northern tongue}}. The region north of the AGN where apparently the northern tongue and the circum-nuclear ring merge is characterised by double-peaked line profiles \citep{galliano02}. As this region is out of the central arcsecond, we postpone the quantitative analysis of it to a subsequent publication \citep{mueller08}. The H$_2$ line profiles of the small scale data studied here do not show complex morphologies and thus they were fitted by one Gaussian component as described in Sect.~\ref{kinematics}.    

\begin{figure}
\plottwo{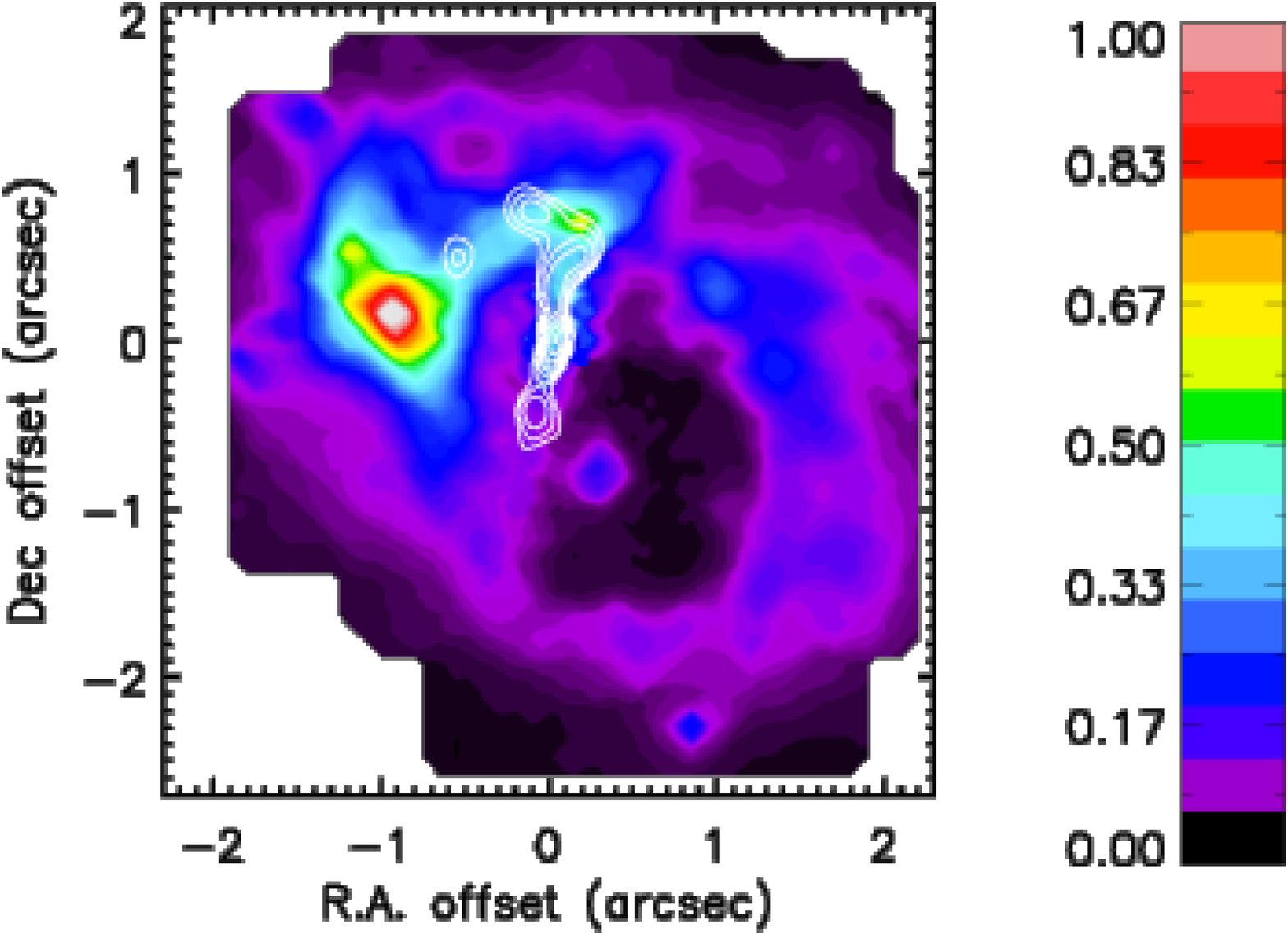}{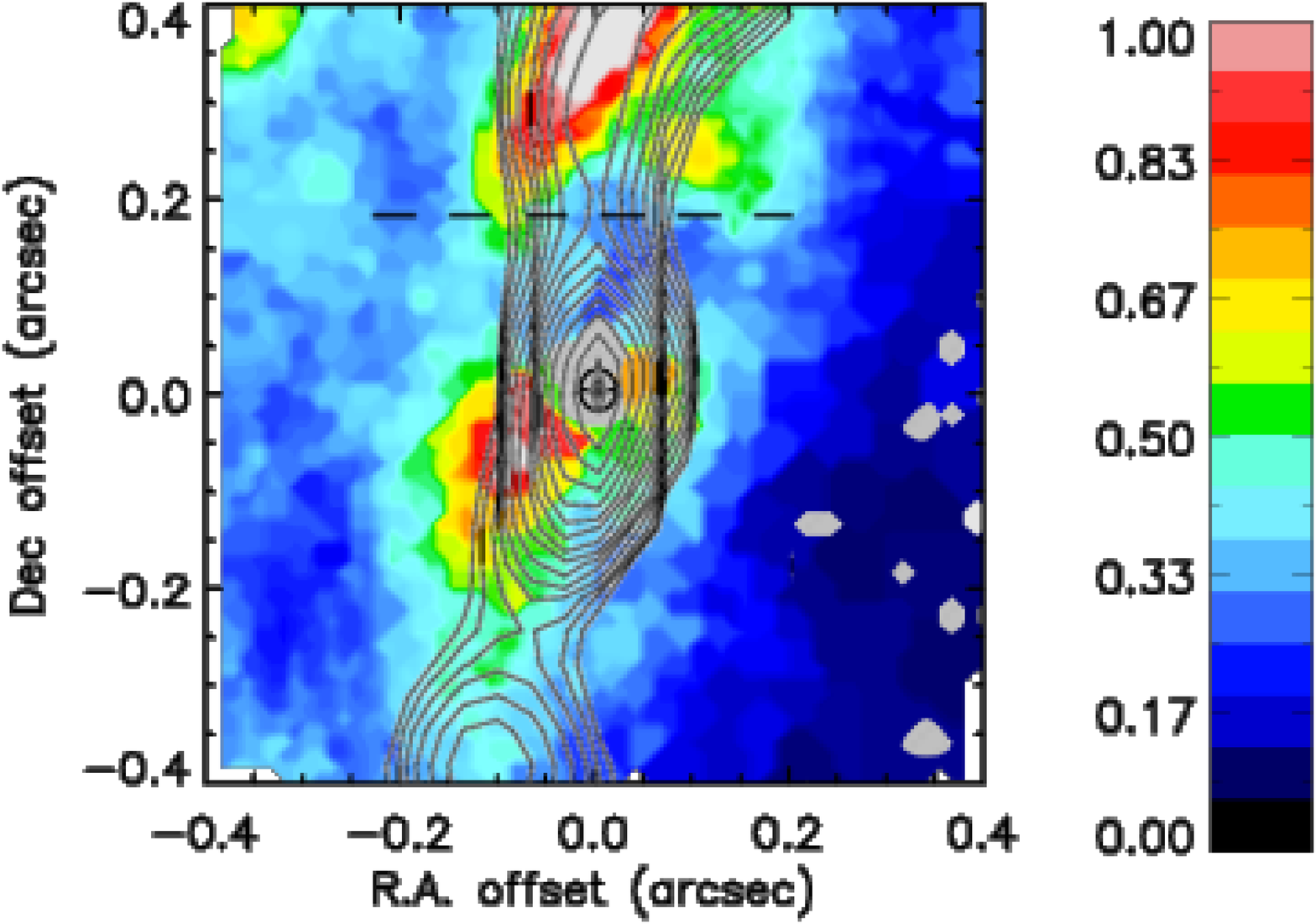}
\caption{Contour plot of the 12.5 $\mu$m deconvolved image of \citet{bock00} overlaid on the SINFONI H$_2$ 1--0 S(1) flux maps. The left panel shows the overlay in the central $4\arcsec\times4\arcsec$ (mid-IR image in white contours) and the right panel in the central $0.8\arcsec\times0.8\arcsec$ of NGC1068. The mid-IR centre is located at the position of the nucleus as defined by the near-IR peak. In the right panel the boundary of the northern tongue is denoted by dashed lines emphazising the good correlation between the two images.\label{fig4}}
\end{figure}

Dust emission at 12 $\mu$m from the northern tongue is present in previous mid-IR observations by \citet{bock00}, \citet{tom06} and \citet{poncelet07}. Overlays of our two H$_2$ flux maps at different spatial scales and the 12.5 $\mu$m deconvolved image of \citet{bock00} are shown in Fig.~\ref{fig4}. The image taken at 12.5 $\mu$m has the highest angular resolution and highest signal to noise
ratio of the mid-IR images from \citet{bock00}. For these reasons we have chosen
to compare this image with our flux maps, in a similar way as these authors did with other datasets. We want to emphasize, however, that all mid-IR images, show qualitatively the same features.
In both panels of Fig.~\ref{fig4} we assumed that the mid-IR peak is located at the position of the nucleus as defined by the near-IR peak. As can be seen in the two overlays, along the northern tongue there exists good correlation between the molecular gas emission and the 12.5 $\mu$m continuum. In the right panel, 
a dashed line indicates the boundary of the northern tongue emphazising this correlation. Also, the alignment gives a fair correlation between the east-west unresolved mid-IR core and the H$_2$ southern tongue. The two overlays show a bend to the east, although the mid-IR bend is more pronounced. The qualitative agreement of the two images indicates that in the central arcsecond the molecular gas and the dust have a similar spatial distribution which is predominantly a north-south linear structure about $1\arcsec$ (70 pc) long which contains two bright components: the southern and northern tongues. 
Our observations support the interpretation given by \citet{tom06} suggesting that the northern tongue could be a passage that transports material to the central engine. We show for the first time, direct evidence from the gas kinematics that this is indeed what is occurring (see Sect.~\ref{kinematics}).

\begin{figure}
\epsscale{.99}
\plotone{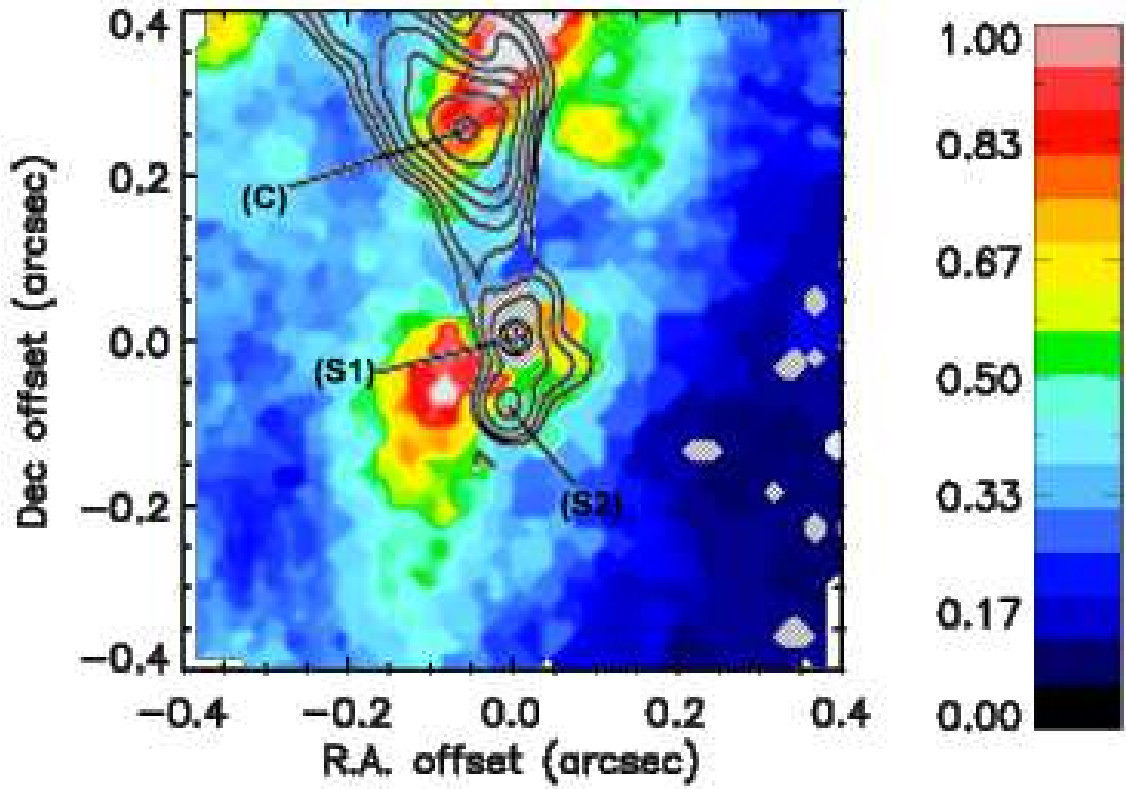}
\caption{Morphology of the H$_2$ 1--0 S(1) emission with the 5 GHz radio continuum overplotted in contours showing the position of the nucleus S1 and the jet-cloud interaction at the position of component C. \label{fig5}}
\end{figure}

Finally, a comparison of the H$_2$ and 5 GHz flux maps is shown in Fig.~\ref{fig5}. Radio continuum imaging at 5 GHz with a resolution of $0.065\arcsec$ revealed a number of structures along the inner part of the radio jet (Gallimore et al. 1996). Associating component S1 with the inner edge of the torus, these authors developed a scenario in which component C arises from a shock interaction between the jet and a dense molecular cloud. Supporting this hypothesis was the bend in the radio jet, the slightly flatter spectral index, and the presence of maser emission. Fig.~\ref{fig5} shows that component C does in fact coincide spatially with the northern tongue of H$_2$ 1--0 S(1) emission. Moreover,
the $\nu=1$ levels are thermalised \citep{mueller08}, indicating that the gas is likely to be rather
dense ($> 10^4$ cm$^{-3}$). Thus the SINFONI data provide direct evidence for the
molecular cloud and hence strongly support the jet-cloud interaction hypothesis.

\subsection{Kinematic evidence for inflow} \label{kinematics}

\subsubsection{Detection of non-circular motions around the nucleus} \label{detection}

\begin{figure}
\plottwo{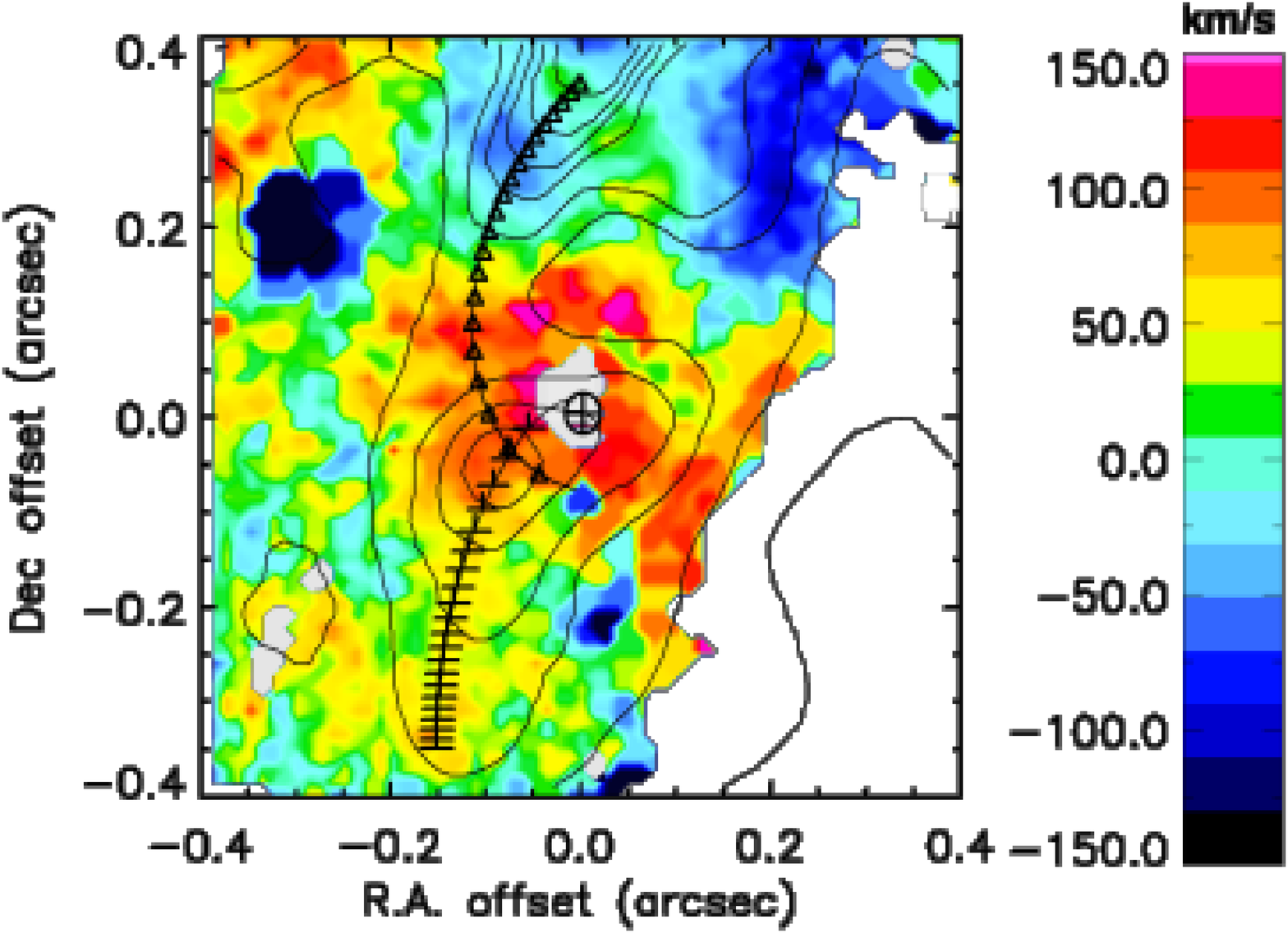}{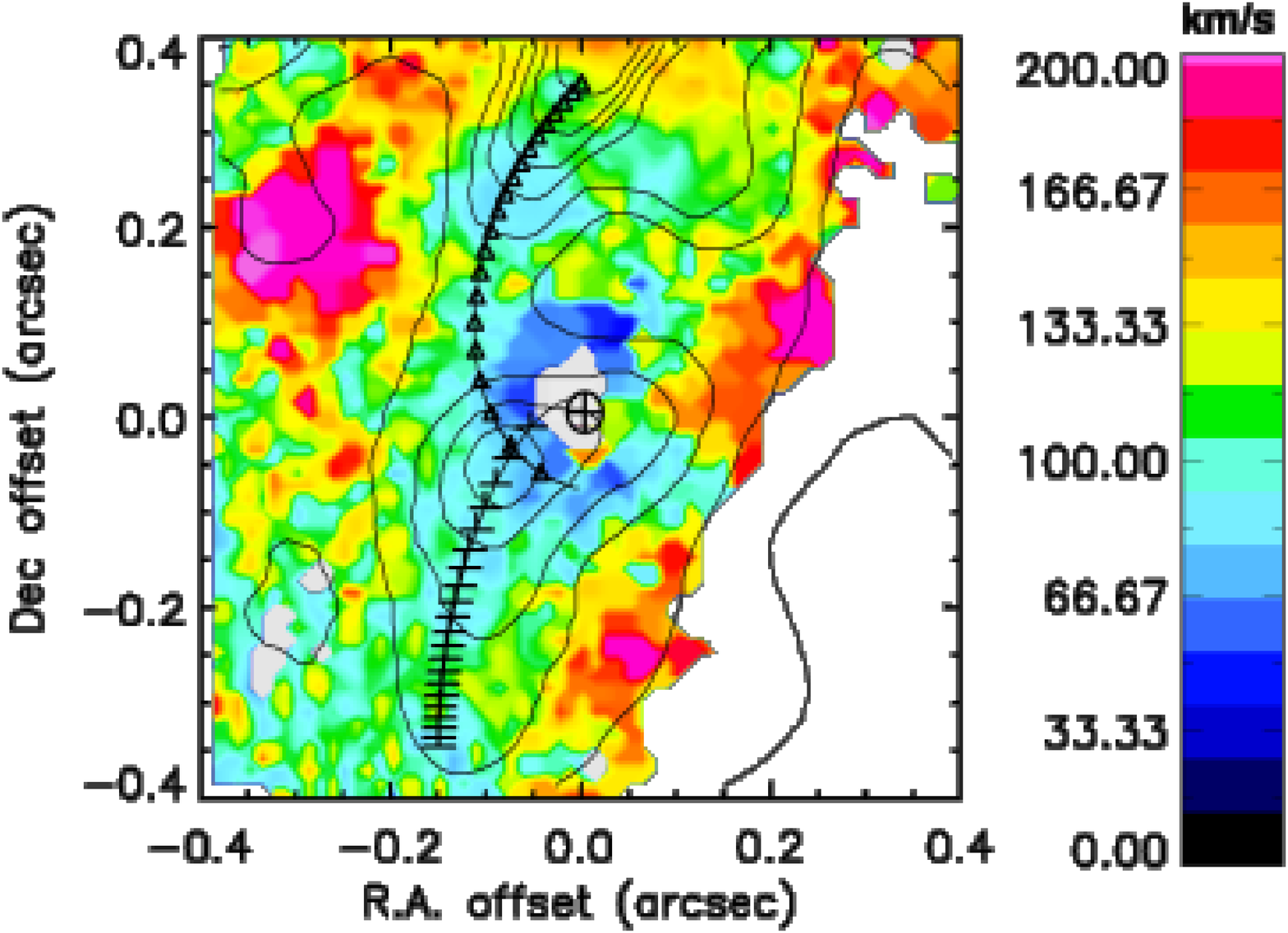}
\caption{Velocity (left) and dispersion (right) maps of the molecular gas extracted from the SINFONI datacube in the central $\pm0.4\arcsec$ of NGC1068. Velocities are measured with respect to the systemic velocity. The dispersion has been corrected for instrumental broadening. A crossed circle indicates in each case the peak of the continuum emission. The maps are binned using Voronoi tesellations \citep{voronoi}. The bins where the line properties could not be extracted were masked out and correspond to the white regions. The rejected pixels in both maps are those with a flux density lower than 10$\%$ of the peak of the central emission shown in Figure~\ref{fig2} and are shown in white in the right part of the fields. The open triangles show the projected trajectory of the northern concentration of gas (the northern tongue, Orbit NT2). The half-crosses show the past trajectory of the gas which is currently located in front of the AGN (the southern tongue, Orbit ST2). \label{fig6}}
\end{figure}

We have extracted gas kinematics in the nuclear region of the galaxy from our integral field data, allowing for the first time a study of the gas motions on scales connecting the outer gas $r\sim1\arcsec$ ($\sim$70 pc) ring to the maser disk at $r=0.015\arcsec$ ($\sim$1 pc). Fig.~\ref{fig6} shows the velocity and dispersion maps of the molecular gas in the central $0.8\arcsec\times 0.8\arcsec$ of the galaxy. 
We have extracted the 2D kinematics by fitting a Gaussian convolved with a spectrally unresolved template profile (an OH sky emission line) to the
continuum-subtracted H$_2$ 1--0 S(1) spectral profile at each spatial pixel in the data cube. We performed a minimisation of the reduced $\chi^2$ in which the parameters of the Gaussian (amplitude, center and width in velocity space) were adjusted until the convolved profile best matched the data. The uncertainties were boot-strapped using
Monte Carlo techniques, assuming that the noise is uncorrelated and the intrinsic profile is well represented by a Gaussian \citep{davies07}. This method allowed us to obtain uncertainties for the velocity and dispersion in the range of $\pm(5-15)$ km s$^{-1}$. The dispersion extracted by this fitting procedure is already corrected for instrumental broadening.  

The velocity field shown in Fig.~\ref{fig6} is quite complex. There is no evidence of a rotating disk around the nucleus. Instead, one can distinguish three main kinematical components: blue-shifted material with almost constant projected velocity $v_z$ of $-25$ km s$^{-1}$ in the north, red-shifted projected velocities between $20-40$ km s$^{-1}$ in the south-east, and a nuclear red-shifted component ($v_z\sim90$ km s$^{-1}$) associated with the southern tongue in Fig.~\ref{fig2}. 
Outside these regions the kinematics are not reliable due to the low strength of the line ($\le 10\%$ of the flux as can be seen in Fig.~\ref{fig2}) and therefore they were masked out in the velocity and dispersion maps. The nuclear red-shifted kinematical component appears to be connected to the south-eastern kinematical component by a ridge of emission that gradually changes its projected velocity from $\sim30$ km s$^{-1}$ at $r=0.4\arcsec$ to $\sim90$ km s$^{-1}$ at $r=0.04\arcsec$ along a PA of $-20\degree$. 
This and the lack of a signature of rotation indicate that the gas must be streaming almost directly towards or outwards from the nucleus rather than orbiting it on circular paths. Thus, any approach to reproduce the observed kinematics by a rotating or warped disk model can be excluded.

\subsubsection{Quantitative modelling} \label{model}

To quantify the kinematics of the region, 
we have modeled the observed velocities as motions of test particles under a gravitational potential comprising a central mass and an extended stellar component. 
 
We define a three-dimensional cartesian coordinate system in the nuclear region of the galaxy with $x-$, $y-$ and $z-$axes representing the Right Ascencion, Declination and Line-of-Sight directions respectively. 
The gravitational potential well of the system is determined by a circum-nuclear mass distribution formed by the sum of a supermassive black hole of $M\mathrm{_{BH}}=1\times10^7$ M$_\sun$ \citep{greenhill96} located at the origin of the system, and a stellar mass density $M_*(r)=1\times10^6\, r$ M$_\sun$ pc$^{-1}$ \citep{davies07, mueller08}. The gas clouds were modelled as test particles -- i.e. they do not have any impact on the potential -- with some initial conditions $x_0\,, y_0\,, z_0\,, v_{x0}\,, v_{y0}$ and $v_{z0}$. The initial $x_0\,, y_0\,$ and $v_{z0}$ components of a test particle are obtained directly from the SINFONI data. Thus, the kinematic model contains three degrees of freedom corresponding to the other three Cartesian components $z_0\,, v_{x0}$ and $v_{y0}$. Notice that $v_{z0}$ can be considered also a variable due to the uncertainity in the observed projected velocities of $\pm10$ km s$^{-1}$. Once the phase space conditions of a cloud are given, the position and velocity of the cloud moving according to the Newtonian laws of motion under the influence of the assumed potential were determined at every time interval $\Delta t=10$ yr over a period of 5 Myr. Therefore, this method creates free unperturbed Keplerian orbits fully defined by the initial values of the Cartesian components of the particle's position and velocity.  

We followed a systematic approach for the determination of the Keplerian orbits of the gas particles. First, an initial position vector $r_0$ is located inside a volume defined by the field of view and several $z_0$ components ranging from $-60$ pc to $60$ pc. This interval was defined as two times the $x-$ or $y-$ range. It is important to point out that this volume delineates the boundaries of the initial position vectors. The resulting orbits are actually contained in a spherical volume of indefinite radius. For each $r_0$ numerous sets of velocity vectors were investigated. The initial $v_{z0}$ component of a particle at any given position in space corresponds basically to the observed projected velocity at that particular spot. The tested values for $v_{x0}$ and $v_{y0}$ at each point in space were dependent on $r_0$ and ranged from $-\sqrt{2GM_\mathrm{{BH}}/r_0}$ to $\sqrt{2GM_\mathrm{{BH}}/r_0}$. This interval was established based on the universal expression for the tangential velocity $v_{tan}$ of any point on the orbit. If the magnitude of the initial tangential velocity vector is smaller than this factor, the motion will be elliptic, if it is larger, the motion will be hyperbolic, and if it is precisely this value, a parabolic orbit will be delineated. In consequence, by considering that the magnitude of its $v_{x0}$ and $v_{y0}$ components range from 0 km s$^{-1}$ to the value given by this factor, all types of orbits are included. Thus, the initial conditions are just a point in the parameter space of any type of orbit of arbitrary eccentricity, size and orientation.   

The final step in the modeling consists of an iterative fitting of orbits to the observed spatial points $(x_{RL}\,, y_{RL})$ in the ridge-line of the intensity map (Fig.~\ref{fig2}) and the projected velocities $(v_z)$. 
We evaluate the goodness of fit by means of an hybrid reduced $\chi^2$ obtained by averaging the reduced $\chi^2$ of each fitted parameter $(x_{RL}\,, y_{RL}\,, v_z)$. We weighted the $\chi^2$ calculation by the square of the uncertainities -- $1/4$ of the spatial resolution for the $x$ and $y$ locations, and for $v_z$ the errors calculated during the extraction of the line properties as discussed in Section~\ref{detection}. If the model is a good approximation to the data, $\chi^2\sim1$. However, the hybrid nature of our $\chi^2$ and the somewhat large uncertainities in the velocity values, could influence the scaling of the reduced $\chi^2$ making its exact value unimportant. In this case, the best-fit will correspond to the orbit presenting the minimum $\chi^2$.

The first attempts to model the kinematics showed that a single Keplerian orbit cannot reproduce the totality of $v_z$ vectors of the observed velocity field. Therefore, we decided to investigate two types of sets of initial conditions as suggested by the observed blue-shifted and red-shifted kinematical components: one for the northern region with $x_0=0\pm5$ pc, $y_0=25\pm5$ pc and $v_{z0}=-25\pm10$ km s$^{-1}$; and one for the southern part with $x_0=-10\pm5$ pc, $y_0=-25\pm5$ pc and $v_{z0}=30\pm10$ km s$^{-1}$. The initial spatial coordinates ($x_0\,,y_0$) of the two sets were selected based on the morphology and kinematics. At first, any good fit to the data should follow the ridge-line. Therefore it is reasonable to have the north and south starting points located on it. Particularly for the northern region, the initial coordinates correspond to the peak of H$_2$ emission in Fig.~\ref{fig2}. Furthermore, they are located in regions where the velocity uncertainties are low. This is once more particularly important for the northern region where the blue-shifted velocities associated morphologically with the northern tongue are well recognizable. Finally, a visual inspection of the velocity field suggests that the motions closer to the nucleus are probably perturbed by other physical processes and could possibly mislead the modeling.

Approximately $5\times10^4$ Keplerian orbits of arbitrary eccentricity, size and orientation were modeled for each region and compared to the observations to determine a best fit. The orbits providing the best approximations to the data at these scales can be further compared with the large scale intensity map shown in Fig.~\ref{fig1} in order to study the connection between the gas in the central arcsecond and the circumnuclear environment. Particularly, we will be able to test the hypothesis of gas streaming from the circumnuclear ring along the linear structure. This hypothesis could motivate a different approach for the determination of the initial conditions ($x_0\,,y_0\,,v_{z0}$). The implicit assumption here is that the gas observed in the central arcsecond (basically the southern and northern tongues) originates from the ring and thus the initial conditions are determined by the crossing points of the linear structure with the ring. However, we decided not to follow this approach because the crossing points of the linear structure with the ring are not clearly determined, especially for the southeastern region where several filaments and knots of gas are connecting the ring to the linear structure (Fig.~\ref{fig1}). In addition, the complex velocity field at these scales contains several kinematical components \citep{mueller08}. This complicates greatly the determination of the initial $v_z$ vectors. Therefore, the approach based on the two sets of initial conditions mentioned in the last paragraph provides a reasonable more general methodology, which can test the hypothesis of gas streaming from the ring along the linear structure, and also determine the exact crossing points of the linear structure with the ring.

\subsubsection{Gas is on highly elliptical/parabolic orbits} \label{elliptical}

\begin{figure}
\plottwo{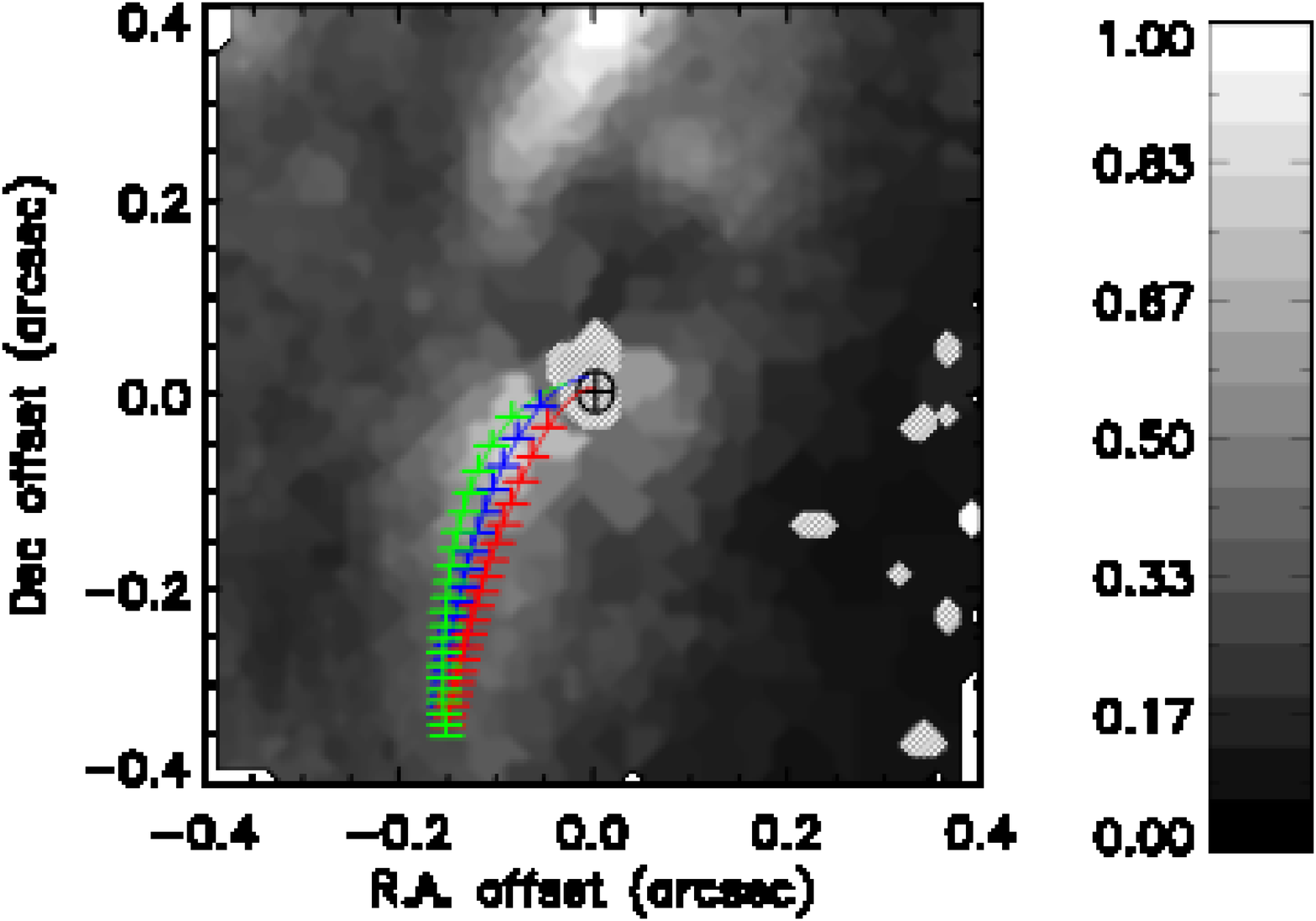}{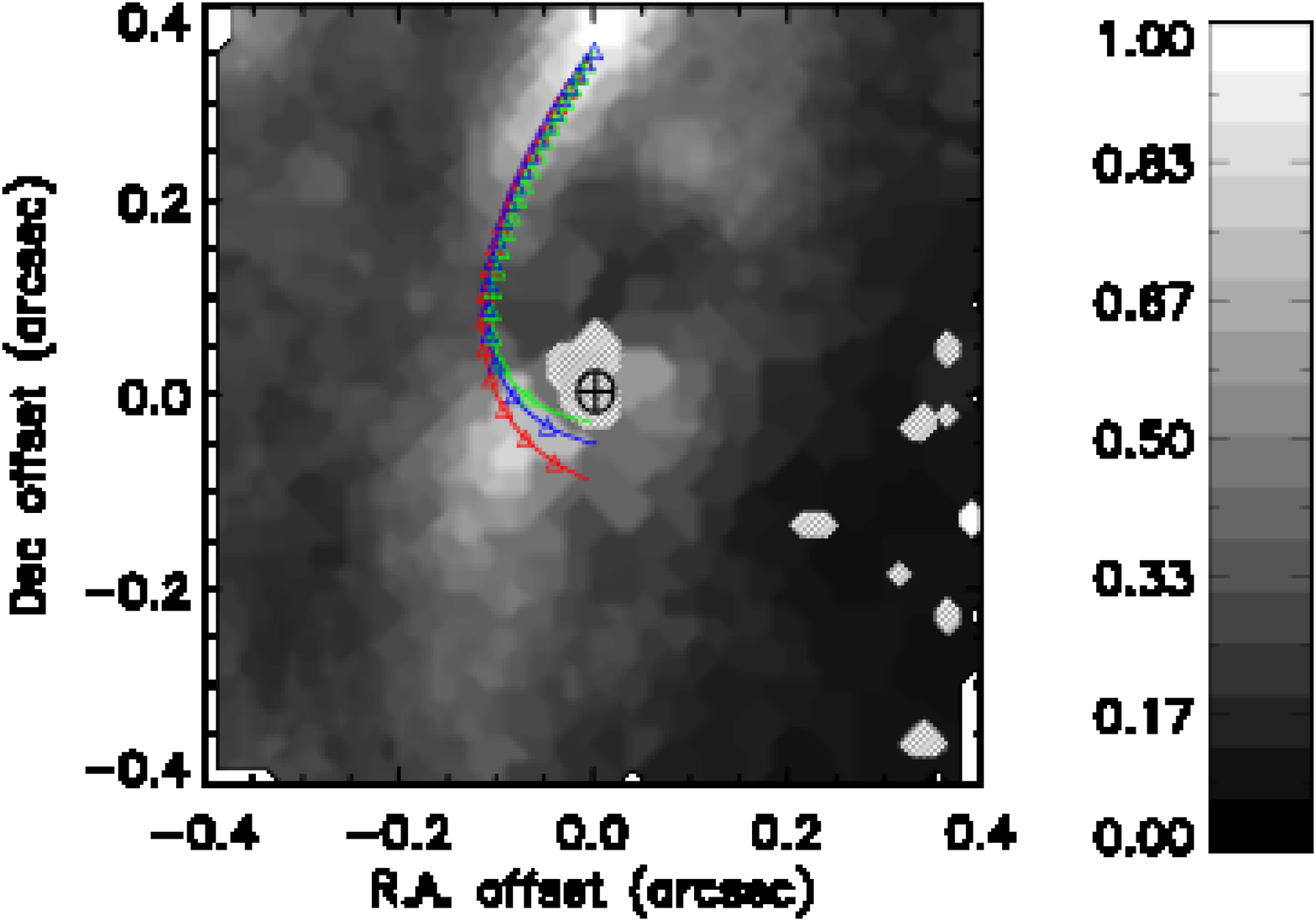}
\caption{\textit{Left:} Trajectories of orbits ST1 (red curve), ST2 (blue curve) and ST3 (green curve) plotted over the intensity map in black-white colour scale. \textit{Right:} Trajectories of orbits NT1 (red curve), NT2 (blue curve) and NT3 (green curve) plotted over the intensity map in black-white colour scale. The peak of the non-stellar continuum is represented by a crossed circle in both panels. The bins where the line properties could not be extracted were masked out and correspond to the diagonal patterned regions in both images. \label{fig7}}
\end{figure}

Based on the reduced $\chi^2$ goodness of fit analysis described in Section~\ref{model}, we found for the northern and southern regions that among the complete set of tested orbits, only highly elliptical/parabolic orbits provide good approximations to the data points $(x_{RL}\,, y_{RL}\,, v_z)$. These orbits are contained in planes that are rotated through an axis coinciding with the $x-$axis (PA of the rotation axis is $90\degree$) and have inclinations between $30-60\degree$. These results confirm the presence of gas streamers in the nucleus of NGC1068. 
The initial conditions and other parameters of three representative orbits of the southern region are shown in Table~\ref{tab1}. The same information for three orbits of the northern region is presented in Table~\ref{tab2}. The trajectories of the six orbits in the projected plane of the galaxy are shown in Fig.~\ref{fig7} plotted over the intensity map. As can be seen in the left panel of this Figure, the southern orbits are spatially associated with the southern tongue, and thus they are identified in Table~\ref{tab1} as orbits/streamers ST1, ST2 and ST3. The northern orbits are clearly associated with the northern tongue (right panel of Fig~\ref{fig7}), and therefore they are designated as orbits/streamers NT1, NT2 and NT3. 

\begin{figure}
\plottwo{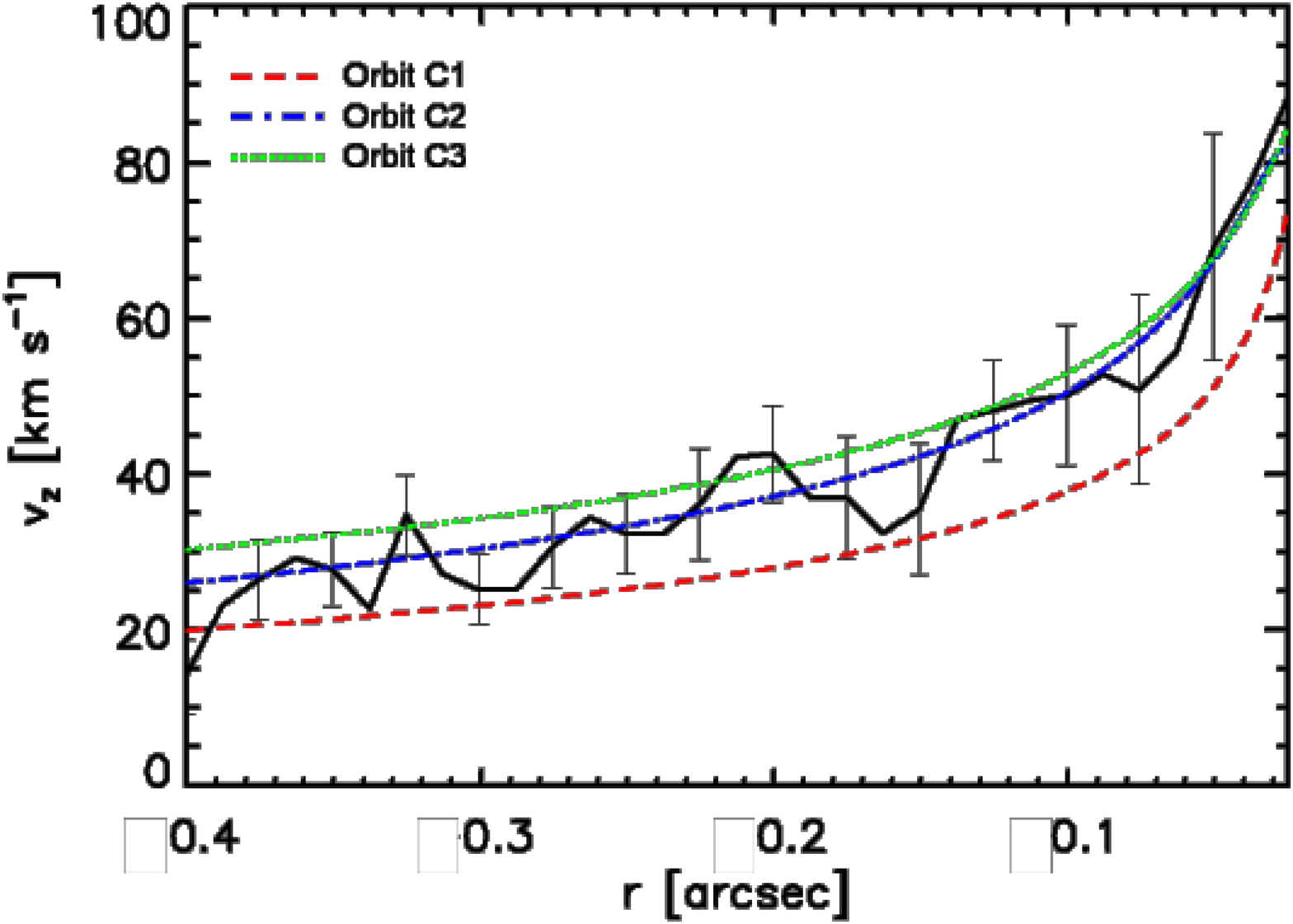}{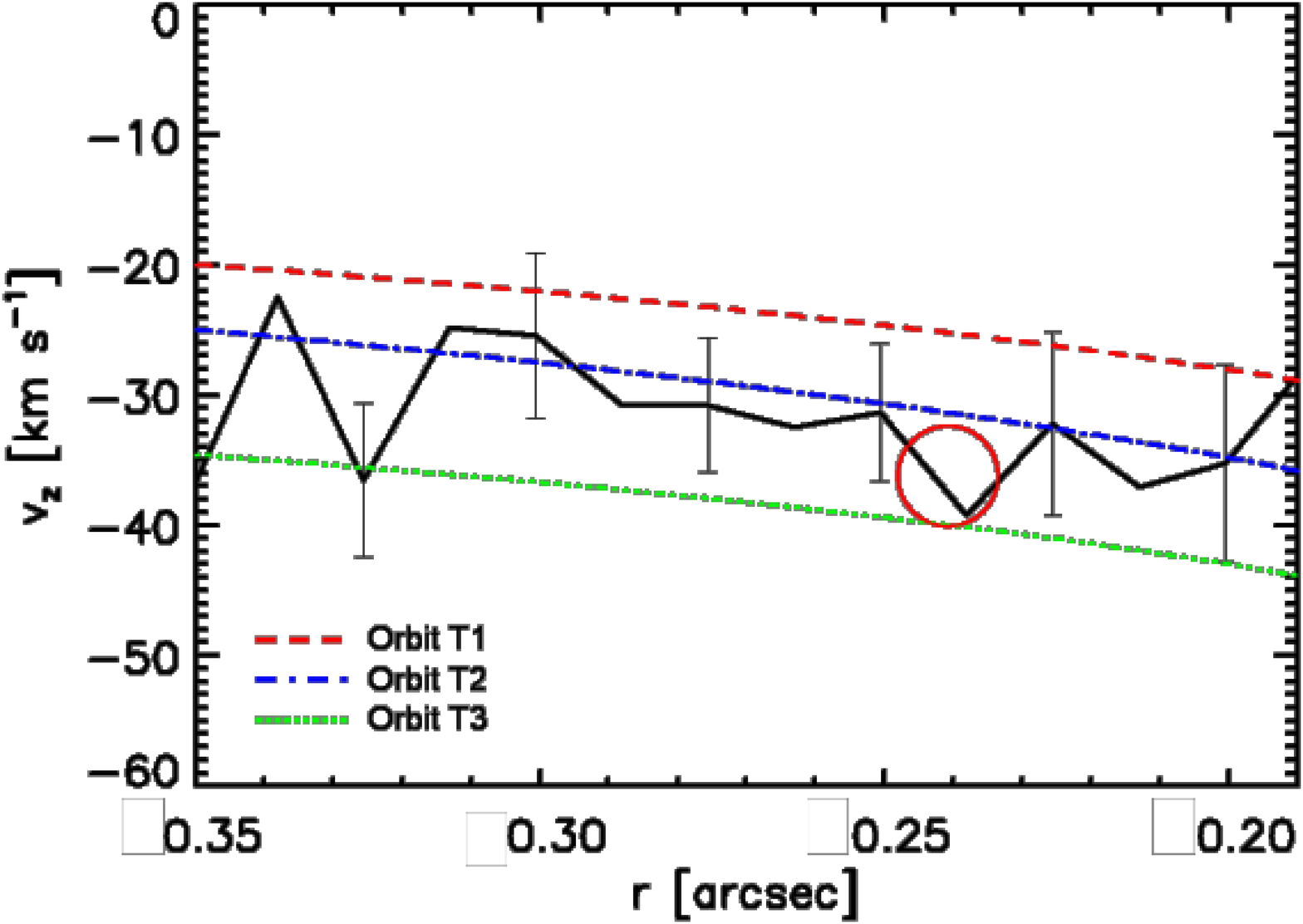}
\caption{Line-of-sight velocities ($v_z$) of the streamers. The left panel shows orbits ST1 (red dashed line), ST2 (blue dashed-dotted line), and ST3 (green dotted line). The right panel shows orbits NT1 (red dashed line), NT2 (blue dashed-dotted line) and NT3 (green dotted line). In the left panel, the black solid line corresponds to the velocity curve extracted from the data along the trajectory of orbit ST2 starting at $r=0.4\arcsec$ until $r=0.015\arcsec$ (1 pc). In the right panel, the black solid line corresponds to the velocity curve extracted from the data along the trajectory of orbit NT2 starting at $r=0.35\arcsec$ until $r=0.2\arcsec$ (14 pc). The jet/cloud interaction in the north takes place at r$\sim0.25$, and is indicated by a red circle in the right panel.  
\label{fig8}}
\end{figure}

The $v_z$ components of the six orbits are presented in Fig.~\ref{fig8}. 
For comparison, velocity curves of the data were extracted along the trajectories of the streamers. As these velocity curves are all similar, only the ones corresponding to orbits ST2 and NT2 are plotted in the left and right panels of Fig.~\ref{fig8} respectively. As can be seen in the two panels of this Figure, there is good agreement between the models and the data within the error bars. The velocity of the gas that is now lying in front of the nucleus corresponds to that of the southern streamer. This is demonstrated in both panels of Fig.~\ref{fig8} and in the left panel of Fig.~\ref{fig6} which shows the trajectories of orbits ST2 and NT2 plotted over the velocity map. It can be seen in this Figure that the southern red-shifted component is connected to the nuclear kinematic component along the streamer trajectory. For the northern region, the projected velocities of the models fit well the data until $r\sim0.17\arcsec$. Within this radius the velocities of the streamer are not recognizable as they are being hidden by the nuclear redshifted kinematical component formed basically by the southern streamer and motions resulting from cloud-cloud collisions at few parsecs from the nucleus.

We evaluate the goodness of fit to the ridge-line and the projected velocities by means of the reduced $\chi^2$ analysis discussed in Section~\ref{model}. The results are shown in Tables~\ref{tab1} and~\ref{tab2}. During the fitting process, it became clear that the somewhat large uncertainities in the velocity values influence the calculation of the reduced $\chi^2$. In consequence, the best-fit corresponds to the orbits presenting the minimum $\chi^2$. This was found for orbits ST2 and NT2 for the southern and northern regions, respectively. The conclusion arising from the orbital fits is that in the central arcsecond of NGC1068 there exist gas streamers flowing 
along the southern and northern tongues in a plane which is rotated around an axis with a PA$=90\degree$ and has an inclination of $45\degree$.

\subsubsection{Streamers flow inwards in the plane of the galaxy}\label{plane}

The orientation of the planes of the orbits discussed in Section~\ref{elliptical} is so that the top half is on the far side and the bottom half towards us. However, the same fits to $(x_{RL}\,, y_{RL}\,, v_z)$ are obtained for orbits located in planes that have the opposite orientation to the one presented in Tables~\ref{tab1} and~\ref{tab2} (negative inclinations). In other words, based only on the kinematic modeling, there are two sets of orbits that we cannot distinguish: those in which the gas is heading towards the nucleus, i.e. the gas is streaming inwards (positive inclinations), and those in which the gas is flowing away from it, i.e. the gas is streaming outwards (negative inclinations). Luckily, our data combined with previous observations provide enough evidence to break this degeneracy.

As demonstrated by several authors, the NE radio jet falls on the near side and its SW counterpart on the far side of the galaxy disk (e.g. Gallimore et al. 1994). The NE jet exhibits a bending point in its way out of the galactic plane caused by a shock interaction between the jet and a dense molecular cloud \citep{gallimore96, tecza01, cecil02, poncelet07}. The direct evidence for the molecular cloud obstructing the way of the jet in our SINFONI data (Fig.~\ref{fig5}) indicates that the orientation of the plane containing the cloud must be compatible with that of the galaxy plane, i.e. the top half is on the far side and the bottom half towards us. In consequence, the gas transport is towards the nucleus, i.e. the gas is streaming inwards.    
 
We can compare this scenario with previous observations of this galaxy. 
According to the "NGC1068 Ringberg standards" \citep{bland97}, the galaxy disk has an inclination of $i=40 \pm3\degree$ and a kinematic major axis close to $90\degree$. A similar configuration was found from the kinematics of the stars observed at $r=3\arcsec$ ($\sim200$ pc) with SINFONI \citep{davies07, mueller08}. These authors found that a rotating disk model with a PA$=84\pm4\degree$ and $i=45 \pm4\degree$ reproduces the stellar kinematics well. At scales down to $r=0.5\arcsec$, the stellar velocity field is compatible with the same kinematic major axis but the inclination is not well constrained. This situation is similar with what we have found for the gas, although our modeling favours an inclination of $45\degree$.       

The orientation is consistent with the strong evidence across the electromagnetic spectrum that the NE radio jet, ionization cones and the extended x-ray emission fall on the near side, and the SW counterparts of all these fall on the far side, and therefore hidden by the galaxy disk \citep{bland97}. 
Finally, the plane orientation of our model is also consistent with the recent H$_2$ observations of \citet{galliano02} who reproduced the observed changes in the line profiles across the 200 pc central region by a two-component kinematical model consisting of a rotating molecular disk inclined $65\degree$ with this orientation plus an outflow.

Thus, the resulting physical model consists of gas streamers located in the plane of the galaxy, and falling inwards from either side of the nucleus. The streamer in all cases approaches the centre (pericenters of a few pc), but as far as we can tell with the present spatial resolution, it does not actually reach the point source (see Tables~\ref{tab1} and~\ref{tab2}). However, cloud-cloud collisions at few parsecs from the nucleus may provide a mechanism through which orbiting gas can lose angular momentum and remain in the inner region, falling eventually to the centre. 
Additionally, it can be seen in the right panel of Fig.~\ref{fig6}, that the streamers constitute a $\sigma$-dip in the whole field and thus are cold, indicating that the gravity dominating the overall flow velocity is stronger than the internal dispersion. The tidal stretching of the material during infall produces the streamers 
now observed as a linear structure. These motions are strong evidence that we are seeing, on scales down to a few parsec, how gas is being driven toward the AGN in NGC1068, and hence how the AGN is being fuelled.

\subsection{Mass of the infalling molecular material} \label{mass}

We now discuss the molecular gas masses of the southern and northern tongues. The lack of robust direct probes of the gas mass complicates the analysis. To mitigate the uncertainties, we have used several independent methods.  

Our first estimate is based on the dynamical mass obtained by \citet{davies07} from the stellar kinematics in the central arcsecond of the galaxy. A dynamical mass of $M_{dyn}=1.3\times10^8$ M$_\odot$ within $r=0.5\arcsec$ (35 pc) is given by these authors. Assuming a conservative 10\% gas fraction \citep{hicks08} and that most of the gas mass within this radius is distributed in equal amounts inside the southern and northern tongues (see Fig.~\ref{fig2}), each of these components would have then a mass of $\sim6\times10^6$ M$_\odot$. This value can be considered as a lower limit due to the small adopted gas fraction, which probably is higher in the nuclear regions of Seyfert 2 galaxies \citep{hicks08}. 

On the other hand, assuming that the gas particles in each of the two tongues had initially a spheroidal geometry, we can estimate a dynamical mass directly from the dispersion velocity from the relation $M_{dyn}=5\sigma^2r/G$ \citep{bender92}. 
Considering a spheroid with radius $R=0.1\arcsec$ (7 pc) based on the approximate FWHM of the southern tongue, and $\sigma=70\,\pm15$ km s$^{-1}$ as measured from the dispersion map along the trajectories of the streamers (Fig.~\ref{fig6}), we find a mass of $6.5\,\pm0.1\times10^7$ M$_\odot$ for each concentration of molecular gas. The obtained value represents an upper limit on the gas mass since this calculation implicity assumes that the gas is gravitational bound and this may not be the case. 

A third independent method to estimate the mass of the molecular gas in the region is 
from the H$_2$ 1--0 S(1) luminosity and an appropriate conversion factor between this quantity and $M_{\mathrm{gas}}$. \citet{mueller06} found from large aperture ($3\arcsec$ or more) measurements of actively star-forming galaxies (including NGC1068) a ratio of 
$L_{\mathrm{1-0S(1)}}/M_{\mathrm{gas}}=2.5\times10^{-4}\, L_\odot/M_\odot$. Although this ratio was obtained for conditions which are probably not met in the nuclear region of the galaxy and depends on fraction of gas on cloud surfaces that can be heated, we apply it simply because there is evidence for vigourous star formation close to the AGN \citep{thatte97, davies07, mueller08} and it can help us to make at least an approximate estimate of the total molecular gas mass. From the total H$_2$ 1--0 S(1) luminosity of the northern and the southern tongue we estimate the total gas mass of each component to be $6\times10^7$ M$_\sun$ and $4\times10^7\, \mathrm{M_\sun}$ respectively. The uncertainties in this approach correspond to the uncertainties in the conversion factor which has a $1\sigma$ statistical uncertainty from the 17 galaxies in their sample of a factor of 2. However, this factor has additionally a systematic uncertainty inherent to the probable overestimated gas masses of the galaxies in the table from \citet{mueller06} obtained using a standard CO--H$_2$ conversion factor (overestimation of factors 2--5). 
This would lead to masses for the nortehrn and the southern tongue as little as $6\times10^6$ M$_\sun$ and $4\times10^6\, \mathrm{M_\sun}$ respectively. A similar conversion factor between cold and warm gas mass is proposed by \citet{dale05} for NGC1068 $M^{\mathrm{warm}}_{\mathrm{H}_2}/M^{\mathrm{cold}}_{\mathrm{H}_2}=$ $1\times 10^{-6}$. After converting the H$_2$ 1--0 S(1) luminosities to warm gas masses \citep{dale05}, we obtain total gas masses which are very consistent with the values calculated by means of the ratio $L_{\mathrm{1-0S(1)}}/M_{\mathrm{gas}}$.  

The X-ray irradiation of molecular clouds by the central X-ray source in NGC1068 has been discussed by \citet{maloney97} and \citet{matt97}, see also \citet{neufeld94}. Maloney et al. found that an attenuating column density of $10^{22}$ cm$^{-2}$ between the central engine and the H$_2$ emitting molecular clouds in the central $r\sim100$ pc accounts for the observed H$_2$ intensity. As the gas density is a function of radius, with higher densities the further in, it is probable that the column density is increased at our scales. \citet{matt97} have shown that the opaque material that obscures our direct view to the central engine of NGC1068 is Compton thick with an attenuating column density of at least $n_H=10^{24}$ cm$^{-2}$. However, this does not imply that the H$_2$ clouds have precisely, or even approximately,  this value. By assuming this and $10^{23 }$ cm$^{-2}$ as the upper and lower limits of the averaged column density in the central $r=0.4\arcsec$, we find that the southern/northern tongue has a gas mass ranging from $2\times10^6$ to $2\times10^7$ M$_\odot$.   

Our last method for determining the gas masses is based on the dust emissivity at mid-IR wavelengths. Recent mid-IR observations from \citet{tom06} inferred a value of $4.6\times 10^5$ M$_\odot$ for the dust mass in the northern tongue. As the total gas mass is expected to be $\sim100$ times greater than the dust mass \citep{cont03, draine03}, a value of a $4.6\times 10^7$ M$_\sun$ can be inferred for the gas mass in this component. Assuming once more that the two gas concentrations have basically the same mass, the southern tongue will have also a gas mass of $\sim4.6\times10^7$ M$_\odot$. 

We can compare the values derived from the five different methods with previous estimates of the gas mass in the nuclear region of NGC1068. From millimeter/sub-millimeter interferometry, \citet{tacconi94} estimated the total mass contained within 1$\arcsec$ of the nucleus as $\sim1.6\times10^8$ M$_\sun$. More recent interferometric observations of the $^{12}$CO(1--0) and $^{12}$CO(2--1) emission by \citet{schinnerer00} give a molecular mass of $\sim5\times10^7$ M$_\odot$ contained in the central 100 pc of the galaxy. In addition, radiative transfer theoretical models from \citet{schart05} predict a dust mass of $8\times 10^4$ M$_\odot$ in the central 70 pc, which converted to gas mass results in a value of $8\times 10^6$ M$_\odot$.   

All estimates are plausibly consistent with each other and with previous approximations suggesting that each molecular gas concentration (the southern or the northern tongue) has a mass within the range $6\times10^6 - 6\times10^7$ M$_\odot$ with a logarithmic mean of $2\times10^7$ M$_\odot$ and a statistical uncertainty of a factor of 3. We will adopt this value for the mass of each component. This is consistent with a 25\% gas fraction of the dynamical mass in the central arcsecond of the galaxy \citep{tacconi94, thatte97, schinnerer00, davies07, mueller08}. Furthermore, it is consistent with the molecular mass estimated from $^{12}$CO(2--1) emission \citep{schinnerer00} on slightly larger scales and an averaged column density in the central $\pm0.4\arcsec$ of $10^{24}$ cm$^{-2}$.

\subsection{Mass accretion rate} \label{accretion}

The morphology and kinematics of the gas are strong evidence that we are witnessing, on scales of a few parsecs, how gas is being driven toward the AGN in NGC1068, and hence how the nucleus is being fuelled. 

The mass accretion rate down to a few parsecs from the AGN can be estimated assuming that material falls into the nucleus through the linear structure. The infalling time scale, defined as the time a gas cloud takes to travel from the initial position to the pericenter, is obtained directly from the modeling and has a value of 1.3 Myr for the northern streamer (see Table~\ref{tab2}). This factor and the total gas mass yield a mass accretion rate at these scales of $\sim15$ M$_\odot$ yr$^{-1}$ with a $1\sigma$ uncertainty of a factor of 3. This value is an upper limit since not all the gas flowing towards the nucleus will actually stay there. A similar value is found for the southern streamer assuming that its mass corresponds to the mass that currently is located in front of the nucleus. From millimeter interferometry, \citet{tacconi94} estimated this influx to be a few M$_\sun$ yr$^{-1}$ from the total mass contained within 1$\arcsec$ of the nucleus ($\sim1.6\times10^8$ M$_\sun$) and radial velocities ($\sim50$ km s$^{-1}$), which is of the same order as our data. 

We can compare this inflow rate on scales of $\sim10$ pc to that onto the black hole itself. The mass accretion rate at scales down to one Schwarzschild radius $R_S$ can be estimated from the mass-to-luminosity conversion efficiency of a black hole $L=\eta\,c^2\mathrm{d}M/\mathrm{d}t$, where $\eta$ is the accretion efficiency with typical values $0.1-0.3$ \citep{eardley75} and $L$ is the bolometric luminosity of the AGN, which for NGC1068 is $\sim8\times10^{44}$ ergs s$^{-1}$ \citep{telesco80}. This yields a mass accretion rate of $0.03-0.09$ M$_\sun$ yr$^{-1}$, indicating that the $\mathrm{d}M/\mathrm{d}t$ is reduced $\approx1000$ times from $r=1$ pc to a few $R_S$. An analogous situation is observed at our own Galactic Center, for which mass accretion rates of $10^{-3...-4}$ M$_\sun$ yr$^{-1}$ at $r=1$ pc are observed \citep{genzel87}, and $\mathrm{d}M/\mathrm{d}t\sim10^{-7...-8}$ M$_\sun$ yr$^{-1}$ at a few $R_S$ from the point source SgrA* are estimated for a bolometric luminosity of the point source of a few 10$^{38}$ ergs s$^{-1}$ \citep{ozernoy96, marrone07}. 
The apparent $\mathrm{d}M/\mathrm{d}t$ is a strong function of radius, with much reduced $\mathrm{d}M/\mathrm{d}t$ closer to the nucleus. Qualitatively this is a natural consequence of inefficient angular momentum transport. 
While this may be a coincidence, we speculate that the gas accretion mechanisms are similar for active and non-active supermassive black holes. Recent simulations of accretion of stellar winds on to Sgr A$^*$ \citep{cuadra06} and supermassive black holes in Seyfert nuclei \citep{schartmann07}, reveal that the cold gas streamed down to scales approximately ten times smaller and settled into a very turbulent disk. This suggests that indeed the processes both, helping and hindering gas inflow, are the same for Seyfert galaxies and the Galactic Center.

\subsection{Obscuration by inflowing gas} \label{obscuration}

We now discuss the implications of the presence of molecular gas in front of the nucleus to the obscuration of the AGN in NGC1068. First, we calculate the gas mass surface density $\Sigma$ of the southern tongue assuming once again a cloud radius of $R=0.1\arcsec$. For a gas mass of $2\times10^7$ M$_\odot$ we obtain $\Sigma=12\times 10^4$ M$_\sun$ pc$^{-2}$, yielding a column density of $n_H = 8\times 10^{24}$ cm$^{-2}$, which is comparable with the values predicted for highly absorbed objects ($n_H \geq 10^{24}$ cm$^{-2}$) and specifically for Compton thick sources such as NGC1068 \citep{bassani99, matt97} and the column densities of clumpy torus models \citep{nenkova02}. This high column density suggests that the nuclear structure is optically thick in the near-IR and hence this gas concentration can be associated with the obscuring material that is hiding the broad line region. In addition, as this value represents an average on scales of $\sim0.2\arcsec$, the true column density of individual smaller clouds must be larger. The fact that a considerable amount of the 2 $\mu$m and 10 $\mu$m emission from the innermost region ($\sim1-2$ pc) and the maser disk radiation is not obscured despite the presence of large gas column densities,  implies that the structure must be a clumpy medium. Line-of-sight attenuation of all these in this case would be a mere consequence of one or more intervening molecular clouds. 

Our observations therefore suggest that the southern tongue can be associated with the obscuring material that is hiding the nucleus but not in the classical picture of a rotating torus. This scenario is mainly ruled out by the kinematics which do not show any type of rotation near the AGN. However, there are several pieces of evidence that support this association. First, the size scale of the nuclear gas is remarkably similar to those of recent mid-IR observations \citep{bock00, tom01, galliano05, tom06, poncelet07} and those of static torus models, in particular the latest clumpy model of \citet{hoenig06}, for which a size of 15$\times$7 pc (diameter) is predicted for the H$_2$ distribution in this galaxy (H\"onig personal communication). There is also a similarity between the PA of the major axis of the core $\sim120 \degree$, which is consistent with that of the line of maser spots \citep{greenhill96} 
and the 300K dust emission \citep{jaffe07}. Furthermore, the estimated gas mass and column density fully agree with previous observations and torus models (see Sect.~\ref{mass}). Hence we interpret this nuclear concentration of gas as a set of infalling clouds, that form the optically thick outerpart of an amorphous clumpy molecular/dusty structure. 
This large scale structure that we have observed will most probably enclose smaller clouds, qualitatively similar to a nested clouds scenario in which a distinct rotating molecular/dusty torus may or may not be present. 
In any case, based on the morphology and kinematics of the gas, we can state that if there is a rotating torus in NGC1068, its outer radius $R_{\mathrm{out}}$ must be smaller than 7 pc and it is encircled by this amorphous molecular/dust obscuration. 

\subsection{Connection of the streamers with the circumnuclear environment} \label{connection}
 
The remaining subject to investigate is the origin of the infalling material. A detailed study of the physical conditions of the NLR and the molecular gas at larger scales is crucial for the understanding of this phenomenon. 
We postpone our analysis of these and other features in the central $3\arcsec\times3\arcsec$ of NGC1068 to a subsequent publication \citep{mueller08}. 
Here, we provide a few remarks on the possible connection of the streamers with the circumnuclear environment.
 
In the central arcsecond, the orbital fit shows that highly elliptical/parabolic orbits in the plane of the galaxy reproduce the ridge and velocities of the gas quite well. A visual inspection of Fig.~\ref{fig1} shows that apparently the central two streamers are part of a linear structure connecting the NNE and SSW to the circumnuclear ring. 
We can test this hypothesis by comparing the complete trajectories of the orbits with the larger scale H$_2$ emission map in Fig.~\ref{fig1}. 
A streamer is consistent with an origin in the ring if its apocenter lies inside the ring. If the apocenter of the orbit lies outside the ring, a ring origin of the streamer is uncertain.  
In order to have a reference of the width of the ring, 
we have superimposed on the intensity map several concentric thin rings of different sizes but having the same axis ratio and position angle as the observed ring in Fig.~\ref{fig1}. Based on width of brightest H$_2$ emission around the ring, this overlay suggests that one can assign a characteristic thickness to the ring of $\sim0.6\arcsec$ ($\sim42$ pc). 

\begin{figure}
\epsscale{.99}
\plotone{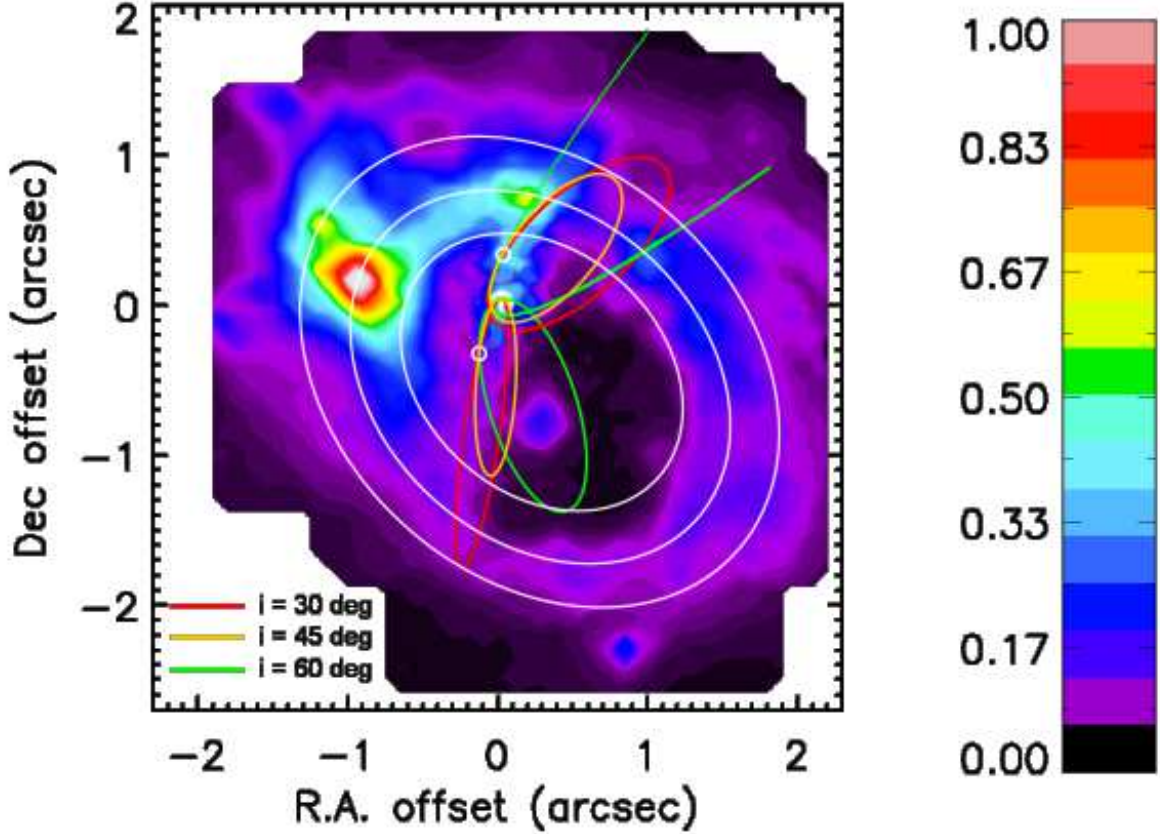}
\caption{Complete trajectories of the six orbits described in Tables~\ref{tab1} and ~\ref{tab2} plotted over the large scale H$_2$ intensity map. Orbits with $30\degree$ inclination (ST1 and NT1) are represented by red curves. Orbits with $45\degree$ inclination (ST2 and NT2) are shown in orange. Orbits with $60\degree$ inclination (ST3 and NT3) are represented by green curves. The location of the AGN as indicated by the peak of the non-stellar continuum is represented by a white crossed circle. The initial spatial coordinates ($x_0$ and $y_0$) of the orbits are indicated by small white circles. The white inner and outer concentric ellipses delinate the thickness of the molecular ring. The central white ellipse is only plotted for reference. 
\label{fig9}}
\end{figure}

The results of this investigation for the six orbits described in Tables~\ref{tab1} and~\ref{tab2} are shown in Fig.~\ref{fig9}. In this Figure, we have plotted the complete trajectories of the orbits over the larger scale H$_2$ flux map presented in Fig.~\ref{fig1}. However, as the gas is streaming inwards, we observe only the part of each orbit which is transporting gas to the nucleus (the eastern part of each orbit). 
Based on the results from the $\chi^2$ goodness of fit analysis, we considered primarily orbits with $45\degree$ inclination. As can be seen in Fig.~\ref{fig9}, orbit NT2 follows the northern part of the linear structure until it apparently merges with the circumnuclear ring. The rest of the streamer's trajectory is fully contained in the ring implying that the streamer is consistent with an origin in the circumnuclear ring. 
In the southern region, orbit ST2 follows the nearly linear extension of the southern tongue until $r\sim0.6\arcsec$. After this point, the linear structure is not distinguishable anymore from the several filaments and knots of gas which appear to be emanating from the ring. 
In this case, as the apocenter of the orbit is contained in the ring (note that the width of the ring can be slightly larger), streamer ST2 is also consistent with an origin in the circumnuclear ring.  

The previous analysis has not only eludicated a plausible connection of the gas streamers with the circumnuclear environment, but also confirmed the results obtained from the $\chi^2$ goodness of fit analysis.  
On one hand, the large scale study suggests that orbit ST3 is not a good fit to the data at these scales as most of its trajectory is actually observed in the empty cavity of the emission map where very little gas is detected. This and the fact that the origin of orbit NT3 is uncertain imply that the gas is with high probability not contained in orbits with inclination angles close to $60\degree$.  
On the other hand, if one asserts that the gas originates in the ring, orbits with 30$\degree$ inclination (ST1 and NT1) provide a moderate fit to the data at these scales. Notice that due to the uncertainty in the thickness of the ring, the origin of streamer NT1 could be associated with the ring. 
In any case, if the streamers originate in the ring, the large scale study clearly favours orbits with inclination angles close to $45\degree$, fully consistent with the results from previous $\chi^2$ analysis (Section~\ref{elliptical}).   

Whether the circumnuclear ring is contained in the plane of the galaxy as the streamers or in another configuration is presently uncertain. 
Assuming a ring origin for the streamers, and that the circumnuclear ring is contained in the plane of the galaxy as suggested by previous authors \citep{schinnerer00, galliano02}, then for material in the ring to approach the center, collisions must remove a significant fraction of the angular momentum. 
There is strong evidence that gas in the ring exhibits significant non-circular motions \citep{galliano02, davies06, mueller08}. This could result in collisions that lead to loss of sufficient angular momentum. 
Such non-rotating clouds would likely collide with orbiting material in the ring at a stationary point in space, the gas torn off the ring would originate at this stationary point, and so should follow a time-independent path. 
The tidal stretching of the material during infall produces the streamers now observed as a linear structure. 
However plausible this scenario may seem, it remains speculative at this point. A better understanding of the physical properties of the circumnuclear ring is required. Thus, as mentioned above, we postpone any interpretation of the ring to a subsequent paper \citep{mueller08}.

\section{Conclusions} \label{conclusions}

We have presented in this paper high resolution SINFONI observations of the molecular gas in the nucleus of NGC1068. The distribution of the H$_2$ 1--0 S(1) emission at a resolution of $0.075\arcsec$ has been resolved. Two bright regions connected by a linear structure extending up to $0.7\arcsec$ north of the nucleus along a PA of $\sim-14\degree$ are distinguished: one lying right in front of the nucleus (the southern tongue) and another one $0.35\arcsec$ north of the center (the northern tongue). The northern tongue correlates with the mid-IR emission and its tip coincides with a knot of radio continuum emission providing direct evidence of the shock interface between the jet and a molecular cloud that has caused the jet direction to change. 
The main results of our analysis on the kinematics of these components are summarized as follows:

\begin{itemize}

\item  Dynamical modeling shows that material is streaming towards the nucleus. The infalling gas is contained on elliptical/parabolic orbits whose orientation is consistent with that of the plane of the galaxy. We interpret this as strong evidence of how gas, on scales of a few parsecs, is fueling the AGN.   

\item The gas transport is from $\sim70$ pc to a few parsecs from the nucleus in the plane of the galaxy. The modeling reveals the existence of two streamers: a northern streamer associated with the northern tongue that passes very close to the nucleus (pericenter of 5 pc), and a southern streamer which lies currently in front of the nucleus associated with the southern tongue, and has a pericenter of $\sim1$ pc.

\item The mass inflow rate $\mathrm{d}M/\mathrm{d}t$ is  $\sim 15$ M$_\odot$ yr$^{-1}$ from scales of 30 pc to a few pc. This is about 1000 times that from a few pc to a few times the Schwarzschild radius $R_S$. A similar change in the mass inflow rate with radius is observed in the Galactic Center. This may be a natural consequence of inefficient angular momentum transport.
 
\item The geometry, kinematics and high column density of the nuclear concentration of molecular gas (the southern tongue) can be explained by a tidally disrupted streamer consisting of a set of infalling clouds that form the optically thick outerpart of an amorphous clumpy molecular/dusty structure.
\end{itemize}

\acknowledgments

The authors are grateful to the staff at the Paranal Observatory for their support during the observations,
and to the entire SINFONI team at MPE and ESO.

\textit{Facilities:} \facility{VLT: Yepun (SINFONI)}.

\clearpage
\begin{landscape}
\begin{table}
\begin{center}
\begin{center}\caption{Orbital parameters of the southern gas streamers\label{tab1}}
\end{center}
\begin{tabular}{cccccccccccccccccc}
\tableline\tableline
\small Orbit & $i$ \tablenotemark{a}& \small PA \tablenotemark{a} & $x_0$ \tablenotemark{b} & $y_0$ \tablenotemark{b} & $z_0$ \tablenotemark{b} & $v_{x0}$ \tablenotemark{c} & $v_{y0}$ \tablenotemark{c} & $v_{z0}$ \tablenotemark{c} &
$x_p$\tablenotemark{d} & $y_p$\tablenotemark{d}& $z_p$\tablenotemark{d} & 
$x_a$\tablenotemark{e} & $y_a$\tablenotemark{e}& $z_a$\tablenotemark{e} &
t\tablenotemark{f} & $\chi^2$\\
\tableline
ST1 & 30 & 90 & -11 & -25 & -14  & 6 & 35 & 20 & 0.1 & 0.42 & 0.24 &
-25 & -121 & -70 & 0.9 & 0.87 \\
ST2 & 45 & 90 & -11 & -25 & -25 & -1 & 25 & 25 & -0.01 & 1.24 & 1.24 &
5 & -70 & -70 & 1 & 0.18 \\
ST3 & 60 & 90 & -11 & -25 & -42 & -1 & 17 & 30 & -0.3 & 1.07 & 1.85 &
28 & -98 & -169 & 1.1 & 0.71 \\
\tableline
\end{tabular}
\tablenotetext{a}{Inclination and position angles are given in ($\degree$)}
\tablenotetext{b}{Initial spatial coordinates in (pc)}
\tablenotetext{c}{Initial velocity vectors in (km s$^{-1}$)}
\tablenotetext{d}{Spatial coordinates of the pericenter in (pc)}
\tablenotetext{e}{Spatial coordinates of the apocenter in (pc)}
\tablenotetext{f}{Infalling time scale in (Myr). This is defined as the time a gas cloud takes to travel from the initial coordinates to the pericenter}
\end{center}
\end{table}
\end{landscape}

\clearpage
\begin{landscape}
\begin{table}
\begin{center}
\begin{center}\caption{Orbital parameters of the northern gas streamers\label{tab2}}
\end{center}
\begin{tabular}{cccccccccccccccccc}
\tableline\tableline
\small Orbit & $i$\tablenotemark{a} & PA\tablenotemark{a}& $x_0$\tablenotemark{b} & $y_0$\tablenotemark{b} & $z_0$\tablenotemark{b} & $v_{x0}$\tablenotemark{c} & $v_{y0}$\tablenotemark{c} & $v_{z0}$\tablenotemark{c} &
$x_p$\tablenotemark{d} & $y_p$\tablenotemark{d}& $z_p$\tablenotemark{d} & 
$x_a$\tablenotemark{e} & $y_a$\tablenotemark{e}& $z_a$\tablenotemark{e} &
t\tablenotemark{f} & $\chi^2$ \\
\tableline
NT1 & 30 & 90 & 0.0 & 25 & 14  & -25 & -35 & -20 & -5 & -3.7 & -2.1 &
82 & 61 & 35 & 1.3 & 1.1 \\
NT2 & 45 & 90 & 0.0 & 25 & 25 & -18 & -25 & -25 & -2.9 & -2.9 & -2.9 &
56 & 54 & 54 & 1.3 & 0.51 \\
NT3 & 60 & 90 & 0.0 & 25 & 42 & -12 & -20 & -35 & -2.4 & -1.8 & -3 &
-- & -- & -- & 1.1 & 2 \\
\tableline
\end{tabular}
\tablenotetext{a}{Inclination and position angles are given in ($\degree$)}
\tablenotetext{b}{Initial spatial coordinates in (pc)}
\tablenotetext{c}{Initial velocity vectors in (km s$^{-1}$)}
\tablenotetext{d}{Spatial coordinates of the pericenter in (pc)}
\tablenotetext{e}{Spatial coordinates of the apocenter in (pc)}
\tablenotetext{f}{Infalling time scale in (Myr). This is defined as the time a gas cloud takes to travel from the initial coordinates to the pericenter}
\end{center}
\end{table}
\end{landscape}

\clearpage

\clearpage

\clearpage

\clearpage

\clearpage

\clearpage
 
\clearpage

\clearpage


\begin{thebibliography}{}

\bibitem[Abuter et al.(2005)]{abuter05} Abuter, R., Schreiber, J., Eisenhauer, F., Ott, T., Horrobin, M., \& Gillesen, S. 2005, NewAR, 50, 398 

\bibitem[Alloin et al.(2001)]{alloin01} Alloin, D., et al., 2001, A\&A, 369, L33


\bibitem[Baker(2000)]{baker00} Baker, A. 2000, Ph.D. thesis, Caltech 


\bibitem[Bassani et al.(1999)]{bassani99} Bassani, L., Dadina, M., Maiolino, R., Salvati, M., Risaliti, G., della Ceca, R., Matt, G., \& Zamorani, G. 1999, ApJS, 121, 473

\bibitem[Bender et al.(1992)]{bender92} Bender, R., Burstein, D., \& Faber, S. M. 1992, ApJ, 399, 462 

\bibitem[Bland-Hawthorn et al.(1997)]{bland97} Bland-Hawthorn, J., Gallimore, J., Tacconi, L., et al. 1997, Ap\&SS, 248, 9

\bibitem[Blietz et al.(1994)]{blietz94} Blietz, M., Cameron, M., Drapatz, S., Genzel, R., Krabbe, A., van der Werf, P.V.D., Sternberg, A., \& Ward, M. 1994, ApJ, 421, 92

\bibitem[Bock et al.(2000)]{bock00} Bock, J. J., et al. 2000, AJ, 120, 2904

\bibitem[Bonnet et al.(2004)]{bonnet04} Bonnet, H., et al. 2004, The ESO Messenger, 117, 17

\bibitem[Cameron et al.(1993)]{cameron93} Cameron, M., et al. 1993, ApJ, 419, 136

\bibitem[Cappellari \& Copin(2003)]{voronoi} Cappellari, M., \& Copin, Y. 2003, MNRAS, 342, 345

\bibitem[Cecil et al. (2002)]{cecil02} 
Cecil, G.,Dopita, M.~A., Groves, B., Wilson, A.~S., Ferruit, P.,
Pécontal, E., Binette, L.\ 2002, ApJ, 568, 627

\bibitem[Contini \& Contini(2003)]{cont03} Contini, M., \& Contini, T. 2003, MNRAS, 342, 299

\bibitem[Cuadra et al.(2006)]{cuadra06} Cuadra, J., Nayakshin, S., Springel, V., \& di Matteo, T. 2006, MNRAS, 366, 358

\bibitem[Dale et al.(2005)]{dale05} Dale, D., Sheth, K., Helou, G., Regan, M., \& H\"uttemeister, S. 2005, AJ, 129, 2197

\bibitem[Das et al.(2006)]{das06} Das, V., Crenshaw, D.M., Kraemer, S.B., \& Deo, R.P. 2006, ApJ, 132, 620

\bibitem[Davies et al.(1998)]{davies98} Davies, R., Sugai, H., \& Ward, M. 1998, MNRAS, 300, 388

\bibitem[Davies et al.(2006)]{davies06} Davies, R., Genzel, R., Tacconi, L., M\"uller S\'anchez, F., \& Sternberg, A. 2006, in Mapping
the Galaxy and Nearby Galaxies, eds. Wada K., Combes F.; arXiv:0610.203

\bibitem[Davies et al.(2007)]{davies07} Davies, R., M\"uller Sanchez, F., Genzel, R., Tacconi, L.J., Hicks, E., Friedrich, S., \& Sternberg, A. 2007, ApJ, 671, 1388

\bibitem[Draine(2003)]{draine03} Draine, B. 2003, ARA\&A, 41, 241

\bibitem[Eardley \& Press(1975)]{eardley75} Eardley, D. M., \& Press, W. H. 1975, ARA\&A, 13, 381

\bibitem[Eisenhauer et al.(2003)]{eisenhau03} Eisenhauer, F., et al. 2003, in Instrument Design and Performance for Optical/Infrared Ground-based Tele-
scopes, eds. Masanori I., Moorwood A., Proc. SPIE, 4841, 1548


\bibitem[Galliano \& Alloin(2002)]{galliano02} Galliano, E., \& Alloin, D. 2002, A\&A, 393, 43

\bibitem[Galliano et al.(2003)]{galliano03} Galliano, E., Alloin, D., Granato, G. L., \& Villar-Mart\'in, M. 2003, A\&A, 412, 615

\bibitem[Galliano et al.(2005)]{galliano05} Galliano, E., Pantin, E., Alloin, D., \& Lagage, P. O. 2005, MNRAS, 363, L1

\bibitem[Gallimore et al.(1996)]{gallimore96} Gallimore, J. F., Baum, S. A., \& O'Dea, C. P. 1996, ApJ, 464, 198

\bibitem[Gallimore et al.(2004)]{gallimore04} Gallimore, J. F., Baum, S. A., \& O`Dea, C. P. 2004, ApJ, 613, 794

\bibitem[Genzel \& Townes(1987)]{genzel87} Genzel, R., \& Townes, C. H. 1987, ARA\&A, 25, 377

\bibitem[Gratadour et al.(2005)]{gratadour05} Gratadour, D., Rouan, D., Boccaletti, A., Riaud, P., \& Cl\'enet, Y. 2005, A\&A, 429, 433

\bibitem[Gratadour et al.(2006)]{gratadour06} Gratadour, D., Rouan, D., Mugnier, L. M., Fusco, T., Cl\'enet, Y., Gendron, E., \& Lacombe, F. 2006, A\&A, 446, 813

\bibitem[Greenhill et al.(1996)]{greenhill96} Greenhill, L. J., Gwinn, C. R., Antonucci, R., \& Barvainis, R. 1996, ApJ, 472, L21


\bibitem[Hicks et al.(2008)]{hicks08} Hicks, E. K. S., et al. 2008, in prep.

\bibitem[H\"onig et al.(2006)]{hoenig06} H\"onig, S., Beckert, T., Ohnaka, K., \& Weigelt, G. 2006, A\&A, 452, 459


\bibitem[Jaffe et al.(2004)]{jaf04} Jaffe, W., et al. 2004, Nature, 429, 47

\bibitem[Jaffe et al.(2007)]{jaffe07} Jaffe, W., Raban, D., R\"ottgering, H., Meisenheimer, K., \& Tristram, K. 2007, ASPC, 373, 439


\bibitem[Maloney (1997)]{maloney97} Maloney, P. 1997, Ap\&SS, 248, 105

\bibitem[Marco et al.(1997)]{marco97} Marco, O., Alloin, D., \& Beuzit, J. L. 1997, A\&A, 320, 399 
	
\bibitem[Marrone et al.(2007)]{marrone07}	Marrone, D. P., Moran, J. M., Zhao, J., \& Rao, R. 2007, ApJ, 654, L57

\bibitem[Mason et al.(2006)]{mason06} Mason, R.E., Geballe, T.R., Packham, C., Levenson, N.A., Elitzur, M., Fisher, R. S., \& Perlman, E. 2006, ApJ, 640, 612

\bibitem[Matt et al.(1997)]{matt97} Matt, G., Guainazzi, M., Frontera, F., et al. 1997, A\&A, 325, L13

\bibitem[Miller \& Antonucci(1983)]{miller83} Miller, J.S., \& Antonucci, R.R.J. 1983, ApJ, 271, L7

\bibitem[M\"uller S\'anchez et al.(2006)]{mueller06}  M\"uller S\'anchez, F., Davies, R.~I., Eisenhauer, F., Tacconi, L.J., Genzel, R., \& Sternberg, A. 2006, A\&A, 454, 481

\bibitem[M\"uller S\'anchez et al.(2008b)]{mueller08}  M\"uller S\'anchez, F., et al. 2008b, in prep.

\bibitem[Nenkova et al.(2002)]{nenkova02} Nenkova, M., Ivezic, Z., \& Elitzur, M. 2002, ApJ, 570, L9

\bibitem[Neufeld et al.(1994)]{neufeld94} Neufeld, D. A., Maloney, P. R., \& Conger, S. 1994, ApJ, 436, L127

\bibitem[Ozernoy \& Genzel(1996)]{ozernoy96} Ozernoy, L., \& Genzel, R. 1996, eds Blitz L., Teuben P., Kluwer Proc. IAU Symposium 169, p.181

\bibitem[Poncelet et al.(2006)]{poncelet06} Poncelet, A., Perrin, G., \& Sol, H. 2006, A\&A, 450, 483

\bibitem[Poncelet et al.(2007)]{poncelet07} Poncelet, A., Doucet, C., Perrin, G., Sol, H., \& Lagage, P. O. 2007, A\&A, 472, 823


\bibitem[Prieto et al.(2005)]{prieto05} Prieto, M. A., Olivier, M., \& Gallimore, J. 2005, MNRAS, 364, L28 

\bibitem[Rotaciuc et al.(1991)]{rotaciuc91} Rotaciuc, V., Krabbe, A., Cameron, M., Drapatz, S., Genzel, R., Sternberg, A., \& Storey, J.W.V. 1991, \apj, 370, L23

\bibitem[Rouan et al.(1998)]{rouan98} Rouan, D., Rigaut, F., Alloin, D., Doyon, R., Lai, O., Crampton, D., Gendron, E., \& Arsenault, R. 1998, A\&A, 339, 687

\bibitem[Rouan et al.(2004)]{rouan04} Rouan, D., et al. 2004, A\&A, 417, L1

\bibitem[Schartmann et al.(2005)]{schart05} Schartmann, M., Meisenheimer, K., Camenzind, M., Wolf, S., \& Henning, T. 2005, A\&A, 437, 861

\bibitem[Schartmann (2007)]{schartmann07} Schartmann, M. 2007, PhD Thesis, MPIA Heidelberg

\bibitem[Schinnerer et al.(2000)]{schinnerer00} Schinnerer, E., Eckart, A., Tacconi, L. J., Genzel, R., \& Downes, D. 2000, ApJ, 533, 850

\bibitem[Tacconi et al.(1994)]{tacconi94} Tacconi, L. J., Genzel, R., Blietz, M., Cameron, M., Harris, A. I., \& Madden, S. 1994, ApJ, 426, L77

\bibitem[Tecza et al.(2001)]{tecza01} Tecza, M., Thatte, N., \& Maiolino, R. 2001, IAUS, 205, 216

\bibitem[Telesco \& Harper(1980)]{telesco80} Telesco, C. M., \& Harper, D. A. 1980, ApJ, 235, 392

\bibitem[Thatte et al.(1997)]{thatte97} Thatte, N., Quirrenbach, A., Genzel, R., Maiolino, R., Tecza, M.\ %
1997, ApJ, 490, 238%

\bibitem[Tomono et al.(2001)]{tom01} Tomono, D., Doi, Y., Usuda, T., \& Nishimura, T. 2001, ApJ, 557, 637

\bibitem[Tomono et al.(2006)]{tom06} Tomono, D., Terada, H., \& Kobayashi, N. 2006, ApJ, 646, 774

\bibitem[Weigelt et al.(2004)]{weigelt04} Weigelt, G., Wittkowski,M., Balega, Y. Y., Beckert, T., Duschl,W. J., Hofmann, K.-H., Men'shchikov, A. B., \& Schertl, D. 2004, A\&A, 425, 77



\bibitem[Wittkowski et al.(2004)]{witt04} Wittkowski, M., Kervella, P., Arsenault, R., Paresce, F., Beckert, T., \& Weigelt, G. 2004, A\&A, 418, L39

\bibitem[Young et al.(1996)]{young96} Young, S., Packham, C., Hough, J. H., \& Efstathiou, A. 1996, MNRAS, 283, L1

\end{thebibliography}
\end{document}